\documentstyle[aps,epsfig,epsf,psfrag]{revtex}

\begin{document}

\title{Spin and charge dynamics of the  ferromagnetic and antiferromagnetic 
two-dimensional half-filled Kondo lattice model.}

\author{S. Capponi and F.~F. Assaad \\
   Institut f\"ur Theoretische Physik III, \\
   Universit\"at Stuttgart, Pfaffenwaldring 57, D-70550 Stuttgart, Germany. }

\maketitle

\begin{abstract}
We present a detailed numerical study of ground state and finite temperature
spin and charge dynamics of the two-dimensional Kondo lattice model with
hopping $t$ and exchange $J$. Our numerical results stem from auxiliary
field quantum Monte Carlo simulations formulated in such a way that the sign
problem is absent at half-band filling thus allowing us to reach lattice sizes
up to $12 \times 12 $.  At $T=0$ and  antiferromagnetic couplings, $J > 0$, 
the competition between the RKKY interaction
and Kondo effect triggers a quantum phase transition between 
antiferromagnetically ordered and magnetically disordered  insulators:
$J_c/t = 1.45 \pm 0.05$. At $J < 0$ the system remains an antiferromagnetically 
ordered insulator and  irrespective of the sign of $J$, the quasiparticle gap scales
as $|J|$.
The dynamical spin structure factor, $S(\vec{q},\omega)$,  evolves 
smoothly from its strong coupling form with spin gap at $ \vec{q} = (\pi,\pi)$
to a spin wave form. 
For $J>0$, the single particle spectral function, $A(\vec{k},\omega)$,
shows a dispersion relation following  that of hybridized bands as obtained 
in the non-interacting periodic
Anderson model. In the ordered phase this feature is supplemented by shadows
thus allowing an interpretation in terms of  coexistence of Kondo screening 
and magnetic ordering.
In contrast, at  $J < 0$ the single particle dispersion relation follows that of 
non-interacting electrons in a staggered external magnetic field.
At finite temperatures spin, $T_S$,   and charge, $T_C$, scales are defined by
locating the maximum in the charge and spin uniform susceptibilities. 
For weak to intermediate couplings, $T_S$  marks the onset of antiferromagnetic
fluctuations - as observed by a growth of the staggered spin susceptibility- 
and follows a $J^2$ law. At strong couplings $T_S$ scales as $J$.
On the other hand $T_C$ scales as $J$ both in the weak and strong coupling
regime. At and slightly below $T_C$ we observe  i) the onset of  screening of
the magnetic impurities, ii) a rise in the resistivity as a function of decreasing 
temperature, 
iii) a dip in the integrated density of states at the Fermi energy and finally 
iv) the occurrence of hybridized bands in $A(\vec{k},\omega)$.
It is shown that in the weak coupling limit, the charge gap 
of order $J$ is formed only at $T_S$ and is hence of magnetic origin. 
The specific heat shows a two peak 
structure. The low temperature peak follows $T_S$ and is hence of magnetic origin.  
Our results are compared to various  mean-field theories.
\\
PACS numbers: 71.27.+a, 71.10.-w, 71.10.Fd  \\ \\ 
\end{abstract}

\section{Introduction}
The Kondo lattice model (KLM) as well  as the  periodic Anderson model 
(PAM) are the prototype Hamiltonians  to describe heavy 
fermion materials~\cite{Lee86} and  Kondo insulators~\cite{Aeppli92}.
The physics under investigation is that of a lattice of magnetic impurities 
embedded in a metallic host. The symmetric PAM reads:
\begin{equation}
\label{PAM}
H_{PAM} =  \sum_{ \vec{k},\sigma }  \varepsilon(\vec{k})
       c^{\dagger}_{\vec{k},\sigma} c_{\vec{k},\sigma}
    - V \sum_{\vec{i},\sigma } 
    \left( c^{\dagger}_{\vec{i},\sigma} f_{\vec{i},\sigma} +
    f^{\dagger}_{\vec{i},\sigma} c_{\vec{i},\sigma}\right)
     + U_f \sum_{\vec{i}} \left( n_{\vec{i},\uparrow}^f -1/2 \right)
	    \left( n_{\vec{i},\downarrow}^f -1/2 \right).
\end{equation}
The unit cell, denoted by $\vec{i}$,  contains an  extended and a
localized orbital. The fermionic operators  $c^{\dagger}_{ \vec{k},\sigma }$
($f^{\dagger}_{ \vec{k},\sigma }$) create electrons on extended (localized)
orbitals with wave vector $\vec{k}$ and $z-$component of spin $\sigma$.
The overlap between extended orbitals generates a conduction band  with
dispersion relation $\varepsilon(\vec{k})$. There is a hybridization 
matrix element, $V$, between both orbitals in the unit-cell  and 
the Coulomb repulsion- modeled by a Hubbard $U_f$- is taken into account
on the localized orbitals. In the limit of large $U_f$, charge fluctuations on
the localized orbitals are suppressed and the PAM maps onto the KLM 
\cite{Schrieffer66}:
\begin{equation}
H_{KLM} =  \sum_{ \vec{k}, \sigma } \varepsilon(\vec{k})
    c^{\dagger}_{\vec{k},\sigma} c_{\vec{k},\sigma}
    + J \sum_{\vec{i}}
    \vec{S}^{c}_{\vec{i}} \cdot \vec{S}^{f}_{\vec{i}}.
\end{equation} 
Here $\vec{S}^{c}_{\vec{i}} = \frac{1}{2} \sum_{s,s'}  c^{\dagger}_{\vec{i},s}
\vec{\sigma}_{s,s'} c_{\vec{i},s'}$, where $\vec{\sigma} $ are the Pauli $s=1/2$
matrices. A similar definition holds for $\vec{S}^{f}_{\vec{i}}$.
A magnetic energy scale $J = 8 V^2/U$  emerges and there is a constraint
of one electron per localized orbital. Although this constraint forbids 
charge fluctuations on the localized orbitals, those fluctuations are
implicitly taken into account leading to the above form and sign 
of the exchange interaction. 
On the other hand, when charge fluctuations on the localized orbitals are absent, 
the exchange interaction follows from Hund's rule and is ferromagnetic.
The ferromagnetic KLM has attracted much attention in conjunction with manganites
\cite{Imada_rev}. In this article we will consider both ferromagnetic and 
antiferromagnetic exchange interactions with emphasis on the
antiferromagnetic case. 

The physics of the single impurity  Anderson  and Kondo models 
at $J/t > 0$
is well understood \cite{Hewson}. In the temperature range $  J < T < U $ 
charge is localized on the $f$-orbital, but the  spin degrees of freedom are
essentially free thus leading to a 
Curie-Weiss law for the impurity spin susceptibility.  Below the Kondo temperature 
$T_K  \propto \varepsilon_f e^{-1/JN(\varepsilon_f)}$ the impurity spin 
is screened by the conduction electrons.  Here, $\varepsilon_f$ is the
Fermi energy and $N(\varepsilon_f)$ the density of states taken 
at the Fermi energy.
The transition from high to 
low temperatures is non-perturbative and corresponds to the Kondo problem
with the known resistivity minimum \cite{Kondo64}
and orthogonality catastrophe  \cite{Anderson67}.  At low temperatures $T_K$
is the only energy scale in the problem.

A lattice of magnetic impurities  introduces new energy scales.
In the spin sector, the  Ruderman-Kittel-Kasuya-Yosida (RKKY) 
interaction~\cite{Kittel63}  couples impurity spins via polarization
of the conduction electrons. This  interaction takes the form 
of a Heisenberg model with exchange $J_{eff}(\vec{q})  
\propto  -J^2 {\rm Re} \chi (\vec{q}, \omega = 0) $ where
$ \chi( \vec{q},\omega)$  corresponds to  the spin  susceptibility of the 
conduction electrons. Since this interaction favors magnetic ordering,
it freezes the impurity spins and hence competes with the Kondo effect.  
By comparing energy scales one expects the RKKY interaction  
(Kondo effect) to dominate at 
weak (strong) couplings. As suggested by Doniach \cite{Doniach77}, 
this leads to  a quantum
phase transition between ordered and disordered magnetic phases. 

As a function of dimension, contrasting results are obtained for the 
PAM and KLM. We first consider the limit of  
large dimensions \cite{Jarrell95,Matsumoto95} and the Gutzwiller 
approximation \cite{Ueda86}. 
The Gutzwiller approximation leads to an  non-interacting PAM with
renormalized hybridization $V$. At half-filling an insulating state is obtained,
with quasiparticle gap $\sim e^{-1/2JN(\varepsilon_f)}$ in the large $U_f$  limit. 
Both the Gutzwiller and  dynamical mean-field approaches yield 
charge and spin  gaps  equal to each other. As a function of temperature, optical
and quasiparticle gaps start appearing at an energy scale  
$\sim e^{-1/2JN(\varepsilon_f)}$ \cite{Jarrell95}.
In the doped phase, the  Luttinger 
volume includes the $f$-electrons, and due to the renormalization of the 
hybridization, the effective mass of charge carriers is enhanced.
The above quoted results stem from calculations for the PAM.
However, similar results are obtained in the 
framework of  the KLM at $J/t << 1$ in the limit of large dimensions 
\cite{Matsumoto95}.  
The above approximations predict an instability to magnetic ordering in the large
$U_f$ or small $J$ limit.  The occurrence of this 
instability has been observed in the 
framework of quantum Monte Carlo (QMC) simulations of the PAM in two dimensions
\cite{Vekic95,Groeber00}.
In the one-dimensional case, a good understanding of the phase diagram of the 
KLM as a function of  electronic density and coupling has been achieved
\cite{Tsunetsugu97_rev,Shibata99}. In particular at half-filling, a spin liquid 
phase is obtained irrespective of the value of $J/t$.
In the weak coupling limit the spin gap follows a Kondo form, whereas the 
charge gap tracks $J$.
 
In this article we  present a detailed numerical study of
ground-state and finite-temperature properties of the half-filled KLM 
in  intermediate dimensions, $d=2$.
Our $T=0$ simulations
are aimed at understanding the competition  and interplay of the Kondo
effect and RKKY interaction.  
Our finite temperature simulations provide  insight into the
temperature evolution of spin and charge degrees of freedom. 

Our main results and structure of the article is as follows.   
Details of the numerical technique are presented in the next section.  
We use a  path integral auxiliary field quantum Monte-Carlo (QMC)
method~\cite{Blankenbecler81}. 
Our approach is based on a simple technical innovation which allows 
to avoid the sign problem at least at half-band filling where
the model is particle-hole symmetric.  Both  finite and zero temperature
versions  of the algorithm are presented. 
In both cases imaginary 
time displaced correlations functions can be computed.
The continuation
to real time is then carried out via the use of the maximum entropy (ME)
method \cite{Jarrell96}. We note that the algorithms 
may be applied irrespective of the sign of $J$.  

In section \ref{T0.sec}  ground state equal time  and dynamical properties of the
ferromagnetic and antiferromagnetic KLM are presented. Our main results include
the following.
i) In the spin sector,  a quantum phase  transition between 
antiferromagnetically ordered and disordered ground  states occurs at 
$J/t = 1.45 \pm 0.05$. The dynamical spin structure factor is analyzed 
across the transition. As a function of decreasing values of $J/t$, the 
spin gap at the antiferromagnetic wave vector closes and the  magnon
spectrum evolves towards  a spin-wave form. This spin wave form persists
for ferromagnetic couplings since in the limit $J/t \rightarrow \infty $  
the model maps onto the $s=1$ antiferromagnetic Heisenberg model. 
Our results at $J/t > 0$ are compared to a bond-operator
mean field theory of the Kondo necklace model.  
ii) In the charge sector, the system remains an insulator. 
To a first approximation, the 
quasiparticle gap tracks $J$ both in the antiferromagnetic and ferromagnetic 
KLM.  
For all values of $J/t > 0 $ the single-particle spectral function 
shows a feature whose dispersion relation follows the one obtained
in the non-interacting PAM. In a mean-field approach, 
this feature results solely from Kondo screening of the magnetic impurities. 
In the magnetically ordered phase, this feature is supplemented 
by shadow bands. Thus and as confirmed by a mean-field approach, 
the spectral function in the ordered phase may
only be understood in terms of the coexistence of Kondo screening
and the RKKY interaction. 
On the other hand, at $J/t < 0$  where Kondo screening 
is absent the single particle dispersion relation follows that of
free electrons  in a external staggered magnetic field.

Section \ref{FT.sec} is devoted to finite temperature properties 
of the KLM. We define charge, $T_C$, as well as spin, $T_S$, scales
from the location of the maximum in the charge and spin susceptibilities. 
In the weak and 
strong coupling limit, the charge scale tracks $J$.  On the other hand
the spin scale - as expected form the energy scale associated to 
the RKKY interaction -
follows a $J^2$ law up to intermediate couplings.
At strong couplings $T_S \propto J$.  
Since $T_C$ corresponds to the energy scale at which a minimum in the 
resistivity is observed, we conclude that it describes the energy
scale at which scattering is enhanced due to the screening of the  impurity
spins. Furthermore a reduction of the integrated density of states  
at the Fermi level is observed at $T_C$.
The spin scale up to intermediate couplings (i.e. $J/t \leq W$ where $W$ corresponds
to the bandwidth) marks the onset of short-range 
antiferromagnetic correlations. This is confirmed by the calculation of the
staggered spin susceptibility which shows a strong increase at $T_S$. 
In the weak coupling limit, it is shown that the quasiparticle gap  of magnitude
$\propto J$ is formed  only at the magnetic energy scale $T_S$ and is thus of
magnetic origin.  In the 
temperature range $ T_S < T < T_C $ hybridized band are seen in the single
particle spectral function with quasiparticle gap lying beyond
our resolution.
Finally, the specific heat is computed and shows a two-peak structure.  
The low energy peak tracks the spin scale and is hence of magnetic origin.

In the last section, we discuss our results as well as links with
experiments. 

\section{Auxiliary field quantum Monte-Carlo algorithm for the
Kondo lattice model}
\label{QMC.sec}
Auxiliary field  QMC  simulations of the KLM as well as the two-impurity Kondo
model  have already been carried out by  Fye and Scalapino as well as by 
Hirsch and Fye~\cite{Fye90,Fye91}. However, their formulation leads to a sign problem even in
the half-filled case where the model is invariant under a particle-hole transformation. 
In this section
we present an alternative formulation of the problem which is free of the sign problem 
in the particle-hole symmetric case. In order to achieve our goal, we take a
detour and 
consider the  Hamiltonian:
\begin{equation}
H =  \sum_{ \vec{k},\sigma }  \varepsilon(\vec{k})
    c^{\dagger}_{\vec{k},\sigma} c^{\phantom{\dagger}}_{\vec{k},\sigma}
    - \frac{J}{4} \sum_{\vec{i}} \left[  \sum_{\sigma}
    c^{\dagger}_{\vec{i},\sigma} f^{\phantom{\dagger}}_{\vec{i},\sigma} +
    f^{\dagger}_{\vec{i},\sigma} c^{\phantom{\dagger}}_{\vec{i},\sigma}
   \right]^2.
\label{Ham1}
\end{equation}
As we will see below, at vanishing chemical potential this Hamiltonian has all 
the properties required to formulate a sign-free auxiliary field QMC algorithm.  
Here, we are interested in ground-state properties  of $H$ which we obtain by 
filtering  out the ground state 
$|\Psi_0 \rangle $ by propagating a trial wave function 
$ |\Psi_T \rangle $ along the imaginary time axis:
\begin{equation}
\frac{\langle \Psi_0 | O |  \Psi_0 \rangle }
           {\langle \Psi_0 |  \Psi_0 \rangle }
            = \lim_{ \Theta \rightarrow \infty }
  \frac{ \langle \Psi_T |e^{-\Theta H } O
        e^{-\Theta H } | \Psi_T \rangle }
     { \langle \Psi_T |e^{-2\Theta H } | \Psi_T \rangle }
\end{equation}
The above equation is valid  provided that $ \langle \Psi_T | \Psi_0 \rangle \neq 0 $ and
$O$ denotes an arbitrary observable.

To see how $H$  relates to $H_{KLM}$ we compute the square in Eq.~(\ref{Ham1}) to obtain:
\begin{equation}
H =  \sum_{ \vec{k}, \sigma } \varepsilon(\vec{k})
    c^{\dagger}_{\vec{k},\sigma} c^{\phantom{\dagger}}_{\vec{k},\sigma}
    + J \sum_{\vec{i}}
    \vec{S}^{c}_{\vec{i}} \cdot \vec{S}^{f}_{\vec{i}} 
  -\frac{J}{4} \sum_{\vec{i},\sigma}
   \left( c^{\dagger}_{\vec{i},\sigma} c^{\dagger}_{\vec{i},-\sigma}
         f_{\vec{i},-\sigma} f_{\vec{i},\sigma}  + {\rm H. c.}
   \right)  
  + \frac{J}{4} \sum_{\vec{i}} \left( n^{c}_{\vec{i}} n^{f}_{\vec{i}} -
                n^{c}_{\vec{i}} - n^{f}_{\vec{i}} \right).
\end{equation}
As apparent, there are only pair-hopping processes between the $f$- and $c$-sites. Thus 
the total number of doubly occupied  and empty $f$-sites is a conserved quantity:
\begin{equation}
[H, \sum_{\vec{i}} (1 - n^{f}_{\vec{i},\uparrow}) (1 - n^{f}_{\vec{i},\downarrow})
 + n^{f}_{\vec{i},\uparrow} n^{f}_{\vec{i},\downarrow} ] = 0.
\end{equation}
If we denote by $Q_n$ the projection onto the Hilbert space 
with $\sum_{\vec{i}} (1 - n^{f}_{\vec{i},\uparrow}) (1 - n^{f}_{\vec{i},\downarrow})
 + n^{f}_{\vec{i},\uparrow} n^{f}_{\vec{i},\downarrow} = n$  then:
\begin{equation}
   H Q_0 = H_{KLM}  + \frac{JN}{4}
\end{equation}
since in the $Q_0$ subspace the  $f$-sites are singly occupied and hence the 
pair-hopping term  vanishes. 
Thus, it suffices to choose  
\begin{equation}
  Q_0 |\Psi_T \rangle  = |\Psi_T \rangle   	
\end{equation}
to ensure that
\begin{equation}
  \frac{ \langle \Psi_T |e^{-\Theta H } O
        e^{-\Theta H } | \Psi_T \rangle }
     { \langle \Psi_T |e^{-2\Theta H } | \Psi_T \rangle } =
 \frac{ \langle \Psi_T |e^{-\Theta H_{KLM} } O
        e^{-\Theta H_{KLM} } | \Psi_T \rangle }
     { \langle \Psi_T |e^{-2\Theta H_{KLM} } | \Psi_T \rangle }.
\end{equation}

It is interesting to note that there is an alternative route to obtain the KLM. 
Instead of projecting  onto the $Q_0$ Hilbert space, we can project onto 
the $Q_N$ 
Hilbert space  by suitably choosing the trial wave function. 
\begin{equation}
H Q_N =  \sum_{ \vec{k}, \sigma } \varepsilon(\vec{k})
    c^{\dagger}_{\vec{k},\sigma} c_{\vec{k},\sigma}
  -\frac{J}{4} \sum_{\vec{i},\sigma}
   \left( c^{\dagger}_{\vec{i},\sigma} c^{\dagger}_{\vec{i},-\sigma}
         f_{\vec{i},-\sigma} f_{\vec{i},\sigma}  + {\rm H. c.}
   \right)  
  + \frac{J}{4} \sum_{\vec{i}} \left( n^{c}_{\vec{i}} n^{f}_{\vec{i}} -
                n^{c}_{\vec{i}} - n^{f}_{\vec{i}} \right).
\end{equation}
Since in the $Q_N$ subspace the $f$-sites are doubly  occupied or empty, 
the exchange term $ \vec{S}^{c}_{\vec{i}} \cdot \vec{S}^{f}_{\vec{i}} $ vanishes. 
To see the relation with the KLM,  we define the spin-1/2 operators:
\begin{equation}
    \tilde{S}^{+,f}_{\vec{i}} = - (-1)^{i_x + i_y} f^{\dagger}_{\vec{i},\uparrow}
	                                           f^{\dagger}_{\vec{i},\downarrow}, \; 
\tilde{S}^{-,f}_{\vec{i}} = - (-1)^{i_x + i_y} f_{\vec{i},\downarrow}
                                                   f_{\vec{i},\uparrow}, \;
\tilde{S}^{z,f}_{\vec{i}} = \frac{1}{2} ( n^f_{\vec{i}} - 1 )
\end{equation}
which operate on the states:  $ | \Uparrow \rangle_{\vec{i},f}  = - (-1)^{i_x + i_y} 
f^{\dagger}_{\vec{i},\uparrow} f^{\dagger}_{\vec{i},\downarrow} | 0 \rangle $ 
and $ | \Downarrow \rangle_{\vec{i},f} = | 0 \rangle  $   as well as the fermion operators:
\begin{equation}
    \tilde{c}^{\dagger}_{\vec{i},\uparrow} =  c^{\dagger}_{\vec{i},\uparrow}, \;
    \tilde{c}^{\dagger}_{\vec{i},\downarrow} = (-1)^{i_x + i_y} c_{\vec{i},\uparrow}.
\end{equation}
With those definitions, 
\begin{equation}
HQ_N = \sum_{ \vec{k}, \sigma } \varepsilon(\vec{k})
    \tilde{c}^{\dagger}_{\vec{k},\sigma} \tilde{c}_{\vec{k},\sigma} 
           + \frac{J}{2} \sum_{\vec{i}}  \left(
    \tilde{S}_{\vec{i}}^{+,c} \tilde{S}_{\vec{i}}^{-,f} + 
       \tilde{S}_{\vec{i}}^{-,c} \tilde{S}_{\vec{i}}^{+,f}  \right) 
          + J \tilde{S}_{\vec{i}}^{z,c} \tilde{S}_{\vec{i}}^{z,f}   + \frac{JN}{4}
\end{equation} 
which is nothing but the KLM. 

\subsection{The basic formalism}
Having shown the relationship between $H$ and $H_{KLM}$ we now discuss some technical
aspects of the QMC evaluation of 
$  \langle \Psi_T |e^{-\Theta H } O
        e^{-\Theta H } | \Psi_T \rangle /
     \langle \Psi_T |e^{-2\Theta H } | \Psi_T \rangle $.
With the use of the Trotter formula we obtain:
\begin{equation}
 \langle \Psi_T |e^{-2\Theta H } | \Psi_T \rangle = 
 \langle \Psi_T |  \prod_{\tau = 1}^{M} e^{\Delta \tau H_t} e^{-\Delta \tau H_J } 
    | \Psi_T \rangle  + {\cal O} (\Delta \tau ^2)
\end{equation}
Here $H_t = -t \sum_{ \langle \vec{i}, \vec{j} \rangle, \sigma } 
c^{\dagger}_{\vec{i},\sigma} c_{\vec{j},\sigma} + {\rm H.c.} $, $H_J = -\frac{J}{4} 
\sum_{\vec{i}} \vec{S}^{c}_{\vec{i}} \cdot \vec{S}^{f}_{\vec{i}} $, and $M \Delta \tau = 2 \Theta$.
Strictly speaking, the systematic error produced by the above Trotter decomposition
should be of order $\Delta \tau$. However, if the  trial wave function as well as 
$H_t$ and $H_J$ are  simultaneously real representable, it can be shown that the prefactor of 
the linear $\Delta  \tau$ error vanishes \cite{Fye86,Assaad94}. 

Since we will ultimately want to integrate out the fermionic degrees of freedom, we carry 
out a Hubbard-Stratonovitch (HS) decomposition of the perfect square term
\cite{Assaad97}:
\begin{eqnarray}
	e^{-\Delta \tau H_J}  & =  & \prod_{\vec{i}} 
          e^{  \Delta \tau J/4 
             \left( \sum_{\sigma} c^{\dagger}_{\vec{i},\sigma} f_{\vec{i},\sigma} + {\rm H.c.} 
             \right)^2 } \nonumber \\
         & = & \prod_{\vec{i}} \left(  \sum_{ l = \pm 1, \pm 2}  \gamma(l) 
e^{ \sqrt{\Delta \tau J/4} 
       \eta(l) \sum_{\sigma} c^{\dagger}_{\vec{i},\sigma} f_{\vec{i},\sigma} + {\rm H.c.} }
		+ {\cal O} (\Delta \tau ^4)       \right), 
\end{eqnarray}
where the fields $\eta$ and $\gamma$ take the values:
\begin{eqnarray}
 \gamma(\pm 1) = 1 + \sqrt{6}/3, \; \; \gamma(\pm 2) = 1 - \sqrt{6}/3
\nonumber \\
 \eta(\pm 1 ) = \pm \sqrt{2 \left(3 - \sqrt{6} \right)},  \; \;
 \eta(\pm 2 ) = \pm \sqrt{2 \left(3 + \sqrt{6} \right)}.
\nonumber  
\end{eqnarray}
As indicated, this transformation is approximate and produces on each time slice
a systematic error proportional to  $ \Delta \tau ^ 4 $. 
This amounts to a net systematic
error of order $ M \Delta \tau ^4 \sim 2 \Theta  \Delta \tau^3 $ which for constant
values of the projection parameter is an order smaller that the error 
produced by the
Trotter decomposition. 

The trial wave function is required to be a Slater determinant
factorizable in the spin indices:
\begin{equation}
        | \Psi_T \rangle =
        | \Psi_T^{\uparrow} \rangle  \otimes
        | \Psi_T^{\downarrow} \rangle \; \;
{\rm with} \; \;  | \Psi_T^{\sigma} \rangle  = \prod_{y=1}^{N_{\sigma}}
\left( \sum_x a^{\dagger}_{x,\sigma} P^{\sigma}_{x,y} \right) | 0 \rangle.
\end{equation}
Here, we have introduced the notation $ x \equiv (\vec{i},n) $ where
$\vec{i}$ denotes the unit cell and $n$ the orbital
(i.e. $ a^{\dagger}_{(\vec{i},1),\sigma} = c^{\dagger}_{\vec{i},\sigma}  $
and  $ a^{\dagger}_{(\vec{i},2),\sigma} = f^{\dagger}_{\vec{i},\sigma}  $).
It is convenient to generate $ | \Psi_T^{\sigma} \rangle $ from a
single particle Hamiltonian  $H_0^{\sigma} = \sum_{x,y} a^{\dagger}_x
\left( h_0^{\sigma} \right)_{x,y} a_y $  which has the trial wave
function as non-degenerate ground state. To  obtain a trial wave function which
satisfies the requirements $ Q_0 | \Psi_T \rangle = | \Psi_T \rangle $ we are 
forced to choose $H_0$ of the form:
\begin{equation}
\label{Trial_N}
	H_0 = \sum_{ \langle \vec{i}, \vec{j} \rangle , \sigma } 
\left( t_{\vec{i}, \vec{j}} c^{\dagger}_{\vec{i},\sigma} c_{\vec{j},\sigma} 
      + {\rm H.c.}  \right)
+   h_z \sum_{\vec{i} } e^{i \vec{Q} \cdot \vec{i} } 
	\left( f^{\dagger}_{\vec{i},\uparrow} f_{\vec{j},\uparrow}  -
               f^{\dagger}_{\vec{i},\downarrow} f_{\vec{j},\downarrow} \right)
\label{H_0}
\end{equation}
which generates a N\'eel state  ($\vec{Q} = (\pi,\pi)$) on the localized orbitals.
To obtain a non-degenerate ground state, we impose the dimerization  
\begin{eqnarray}
t_{ \vec{i},\vec{i} + \vec{a}_x }  =  
\left\{
\begin{array}{c}
-t( 1+\delta)  \; \; {\rm if} \; \; i_x  = 2n+1  \\
-t( 1-\delta)  \; \; {\rm if} \; \; i_x  = 2n
\end{array}
\right.,  \; \; \;
t_{ \vec{i},\vec{i} + \vec{a}_y }  =   -t(1+\delta)
\end{eqnarray}
with $\delta << t $.

We are now in a position to integrate out the fermionic degrees 
of freedom to obtain:
\begin{eqnarray}
     \langle \Psi_T | e^{-2 \Theta H} | \Psi_T \rangle =
 \sum_{ \{ l \} }   \left(   \prod_{\vec{i},\tau} \gamma(l_{\vec{i},\tau}) \right) 
    \prod_{\sigma}  \det \left( P^{\sigma \dagger}
    \prod_{\tau = 1}^{M} e^{- \Delta \tau \hat{T} } e^{\hat{J} (\tau) }
         P^{\sigma} \right), 
\end{eqnarray}
where the matrices $\hat{T}$ and $\hat{J}(\tau) $ are defined via:
\begin{eqnarray}
   H_t  = & & \sum_{\vec{k},\sigma} \epsilon(\vec{k}) 
                c^{\dagger}_{\vec{k},\sigma} c_{\vec{k},\sigma} 
       =    \sum_{x,y, \sigma} a^{\dagger}_{x,\sigma} \hat{T}_{x,y}
                                   a_{y,\sigma}  \nonumber \\
    & &\sum_{x,y, \sigma} a^{\dagger}_{x,\sigma} \hat{J}(\tau)_{x,y}
                                     a_{y,\sigma}.
 =  \sqrt{\Delta \tau J/4}  \sum_{\vec{i}, \sigma}
   \eta(l_{\vec{i},\tau}) \left( c^{\dagger}_{\vec{i},\sigma} f_{\vec{i},\sigma} 
+ {\rm H.c.} \right) 
\end{eqnarray}
The HS field $l$ has acquired a space, $\vec{i} $, and time, $\tau$, index. 

The basic ingredients to compute observables are equal-time Green 
functions. They are given by:
\begin{eqnarray}
& &  \frac{\langle \Psi_T |e^{-\Theta H } a_{x,\sigma} a^{\dagger}_{y,\sigma}
    e^{-\Theta H } | \Psi_T \rangle } 
{ \langle \Psi_T |e^{-2\Theta H } | \Psi_T \rangle }   =
 \sum_{  \{ l \} } {\rm Pr} ( l ) 
\langle \langle a_{x,\sigma} a^{\dagger}_{y,\sigma} \rangle \rangle ( l ) 
\; \; {\rm with} \nonumber \\
& & \langle \langle a_{x,\sigma} a^{\dagger}_{y,\sigma} \rangle \rangle ( l ) =
   \left( 1 - U^{>}_{\sigma,l} 
       \left(U^{<}_{\sigma,l} U^{>}_{\sigma,l} \right)^{-1} 
       U^{<}_{\sigma,l} \right)_{x,y}, \nonumber \\
& & U^{>}_{\sigma,l}  =  \prod_{\tau = 1}^{M/2} 
                e^{- \Delta \tau \hat{T} } e^{\hat{J} (\tau) } P^{\sigma} \; \; \;
U^{<}_{\sigma,l}  =  P^{\sigma \dagger} \prod_{\tau = M}^{M/2 + 1} 
                e^{- \Delta \tau \hat{T} } e^{\hat{J} (\tau) }, \; \; {\rm and} 
\nonumber \\
 & & {\rm Pr} ( l )  = \frac{ \left( \prod_{\vec{i},\tau} \gamma(l_{\vec{i},\tau}) \right) 
               \prod_{\sigma}  
    \det \left(  U^{<}_{\sigma,l} U^{>}_{\sigma,l}  \right) }
{ \sum_{\{ l \} } \left(   \prod_{\vec{i},\tau} \gamma(l_{\vec{i},\tau}) \right)
            \prod_{\sigma} 
       \det \left(  U^{<}_{\sigma,l} U^{>}_{\sigma,l}  \right) }. 
\end{eqnarray}
Since, for a given set of HS fields, we are solving a free electron problem 
interacting with an external field a Wick theorem applies. Hence from the 
knowledge of the the single particle Green function at fixed  HS configuration 
we may evaluate all observables. Imgaginary time displaced correlation functions
may equally be calculated \cite{Assaad96a,Feldbach00}.

We are left with the summation over the HS fields which we will carry out with 
Monte-Carlo  methods. In order to do so  without further complication, we
have to be able to interpret ${\rm Pr} ( l )$ as a probability distribution.  
This is possible only provided that $ {\rm Pr} ( l ) \geq 0 $ for all HS 
configurations. In the particle-hole symmetric case the above statement is valid. Starting
from the identity: 
\begin{equation}
 \det \left(  U^{<}_{\uparrow,l} U^{>}_{\uparrow,l} \right)  =
    \lim_{\beta \rightarrow \infty} 
\frac{ 
     {\rm Tr}  \left(
     e^{-\beta H_0^{\uparrow} }  \prod_{\tau = 1}^{M} e^{-\Delta \tau H_t^{\uparrow}}
     e^{H_J^{\uparrow}(\tau) } \right) } 
            { {\rm Tr}  \left( e^{-\beta H_0^{\uparrow} } \right) }
\end{equation}
we can carry out a particle-hole transformation:
\begin{equation}
\label{PH}
	 c^{\dagger}_{\vec{i}, \uparrow} \rightarrow (-1)^{i_x + i_y} 
       c_{\vec{i}, \downarrow}  \; \; {\rm and } \; \; 
         f^{\dagger}_{\vec{i}, \uparrow} \rightarrow -(-1)^{i_x + i_y} 
       f_{\vec{i}, \downarrow}.
\end{equation}
Here, 
$ H_t^{\sigma} = \sum_{x,y} a^{\dagger}_{x,\sigma} \hat{T}_{x,y}  a_{y,\sigma} $
and 
$ H_J^{\sigma}(\tau)=\sum_{x,y} a^{\dagger}_{x,\sigma} \hat{J}(\tau)_{x,y} a_{y,\sigma}$.
Since Eq. (\ref{PH}) corresponds to a canonical transformation, 
the trace remains invariant and 
$ H_0^{\uparrow}$, $ H_t^{\uparrow}$ as well as  $ H_J^{\uparrow}(\tau) $ 
map onto $ H_0^{\downarrow}$, $ H_t^{\downarrow}$ and $ H_J^{\downarrow}(\tau) $
respectively.  Thus we have shown that:
$ \det \left(  U^{<}_{\uparrow,l} U^{>}_{\uparrow,l} \right)
      = \det \left(  U^{<}_{\downarrow,l} U^{>}_{\downarrow,l} \right) $
which leads to   $ {\rm Pr} ( l ) \geq 0 $ for all values of the HS fields. 
Away from half-filling (which would correspond to adding a chemical potential term 
in $H_0$),  particle hole-symmetry is broken and $ {\rm Pr} ( l ) $ may become 
negative. This leads to the well known sign-problem~\footnote{ It is clear that by choosing  $H_0^{\uparrow} = H_0^{\downarrow}$  
thus leading to $P^{\uparrow} = P^{\downarrow}$ would produce positive  values 
of ${\rm Pr}(l)$ for all HS configurations and irrespective of particle-hole 
symmetry. This stands in analogy to the  absence of sign-problem in the attractive
Hubbard model. However, this choice of the trial wave function is incompatible
with the requirement $Q_0 |\Psi_T \rangle = |\Psi_T \rangle $. }.

For the Monte-Carlo sampling of the probability distribution $Pr(l)$, we adopt a 
sequential single spin-flip algorithm. The details of the upgrading procedure
as well as of the numerical stabilization  of the code are similar to those used
for auxiliary field QMC simulations of the Hubbard model \cite{Loh92}.

\subsection{Optimizing the algorithm}

The above straightforward approach for the QMC simulation of $H$ turns out 
to  be numerically inefficient. 
The major reason for this stems from the choice of the trial wave function. 
 The coupled 
constraints i) $Q_0 |\Psi_T \rangle = |\Psi_T \rangle$ and  ii) $|\Psi_T \rangle $ 
is a Slater 
determinant factorizable in the spin indices   make it impossible to choose
a spin-singlet trial wave function (the trial wave function generated
by the single particle Hamiltonian $H_0$ of Eq.~(\ref{H_0}) orders the 
$f-$electrons in a N\'eel states which is not a spin singlet). Since 
we know that the ground state  of the KLM on a finite-size system is a spin 
singlet~\cite{ShenSQ96,Tsunetsugu97a}, we have to filter
out all the spin excited states from the trial wave function to obtain 
the ground state. This is certainly not a problem when we are investigating the
physics of a problem with a large spin-gap as is the case in the limit $J/t >> 1$. 
However, in the limit of small $J/t $  long-range magnetic order is present 
and hence one expects finite-size spin-gap to scale as $v_s/L$ where $v_s$  is  the 
spin velocity and  $L$ the linear size of the system. In this case,  starting 
with a spin-singlet trial wave function   is important to obtain reliable 
convergence~\cite{Assaad96a}.

In order to circumvent the above problem,   we relax the constraint 
$Q_0 |\Psi_T \rangle = |\Psi_T \rangle$  and add a Hubbard term  for the $f$-sites 
to the Hamiltonian.
\begin{equation}
H =  \sum_{ \vec{k},\sigma }  \varepsilon(\vec{k})
    c^{\dagger}_{\vec{k},\sigma} c_{\vec{k},\sigma}
    - \frac{J}{4} \sum_{\vec{i}} \left[  \sum_{\sigma}
    c^{\dagger}_{\vec{i},\sigma} f_{\vec{i},\sigma} +
    f^{\dagger}_{\vec{i},\sigma} c_{\vec{i},\sigma}
   \right]^2  + U_f \sum_{\vec{i}} ( n^f_{\vec{i},\uparrow} - 1/2 ) 
                                   ( n^f_{\vec{i},\downarrow} - 1/2 ).
\label{Ham2}
\end{equation}
This Hamiltonian is again block diagonal in the $Q_n$ subspaces. During the imaginary
time propagation, the components $ Q_n |\Psi_T \rangle $ of the trial wave function
will be suppressed by a factor $  e^{-\Theta U_f n /2} $ in comparison to the component
$ Q_0 |\Psi_T \rangle $.  

The usual procedure to incorporate the Hubbard term in the QMC simulation
relies on Hirsch's  HS transformation \cite{Hirsch83}:
\begin{eqnarray}
\label{HS1}
 \exp \left( - \Delta \tau  U \sum_{\vec{i}}
\left( n^f_{\vec{i},\uparrow } - \frac{1}{2} \right)
\left( n^f_{\vec{i},\downarrow } - \frac{1}{2} \right) \right)  & &
  \\
 = \tilde{C} \sum_{s_1, \dots, s_N  = \pm 1 } & &
  \exp \left(  \tilde{\alpha} \sum_{\vec{i}}  s_{\vec{i}}
\left( n^f_{\vec{i},\uparrow}  -n^f_{\vec{i},\downarrow } \right) \right).
\nonumber
\end{eqnarray}
where $\cosh(\tilde {\alpha})   = \exp \left( \Delta \tau U / 2  \right)$.
As apparent from the above equation, for a fixed set of HS
fields, $ s_1 \dots s_N$,  $SU(2)$-spin symmetry
is broken. Clearly $SU(2)$ spin symmetry is restored after
summation over the HS fields

Alternatively,   one may consider \cite{Hirsch83}
\begin{eqnarray}
\label{HS2}
 \exp \left( - \Delta \tau  U \sum_{\vec{i}}
\left( n_{\vec{i},\uparrow } - \frac{1}{2} \right)
\left( n_{\vec{i},\downarrow } - \frac{1}{2} \right) \right) & &
  \\
 =  C \sum_{s_1, \dots, s_N  = \pm 1 } & &
   \exp \left( i  \alpha \sum_{\vec{i}}  s_{\vec{i}}
\left( n_{\vec{i},\uparrow} + n_{\vec{i},\downarrow }  - 1 \right) \right).
\nonumber
\end{eqnarray}
where  $ \cos(\alpha)   = \exp \left( - \Delta \tau U / 2  \right)$
and $C  = \exp\left( \Delta  \tau U N / 4 \right )/2^N $.
With this choice of the HS  transformation
$SU(2)$ spin invariance is retained for any given HS configuration.   Even
taking into account the overhead of working with complex numbers, one of the
authors 
has argued \cite{Assaad98b} that this  choice of HS transformation produces a
more efficient code.  
 
Having relaxed the condition $Q_0 | \Psi_T  \rangle = | \Psi_T  \rangle $  we are now
free to choose a spin singlet trial wave function which we generate from:
\begin{equation}
\label{Trial_S0}
H_0 =  \sum_{ \vec{k},\sigma }  \varepsilon(\vec{k})
       c^{\dagger}_{\vec{k},\sigma} c_{\vec{k},\sigma}
    - \frac{J}{4} \sum_{\vec{i},\sigma } 
    (c^{\dagger}_{\vec{i},\sigma} f_{\vec{i},\sigma} +
    f^{\dagger}_{\vec{i},\sigma} c_{\vec{i},\sigma})
\end{equation}
which is nothing but  the non-interacting PAM with
hybridization $V=J/4$. The ground state at half-filling is clearly a 
spin singlet.  With this choice of the trial wave function, and  the 
Hubbard-Stratonovitch transformation of Eq.~(\ref{HS2}) the particle-hole
transformation of  Eq.~(\ref{PH}) maps 
$ \det \left(  U^{<}_{\uparrow,l,s} U^{>}_{\uparrow,l,s} \right)$ on
$ \overline{\det \left(  U^{<}_{\downarrow,l,s} 
         U^{>}_{\downarrow,l,s} \right)}$. 
Hence, no sign
problem occurs at half-filling.

\begin{figure}
\epsfxsize=9.0cm
\epsfysize=6.0cm
\begin{center}
\includegraphics[width=9cm,height=6cm]{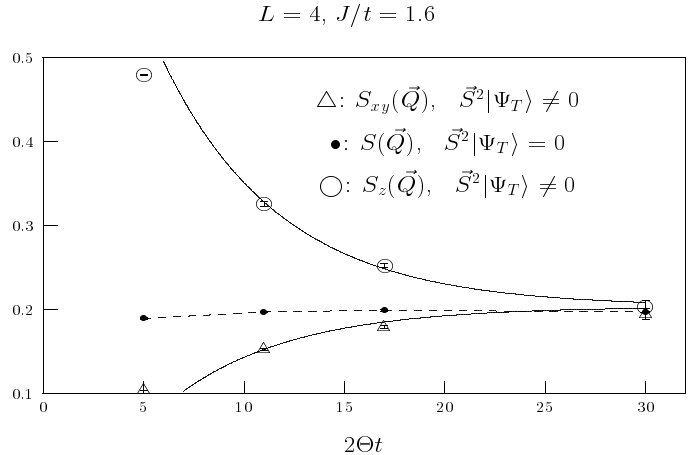}
\end{center}
\caption[]
{\noindent
\label{Conv.fig}  Spin-spin correlations as a function of the projection
parameter $\Theta$. Here,
$S(\vec{Q}) = \frac{4}{3} \langle \vec{S}^f(\vec{Q}) \cdot \vec{S}^f(\vec{-Q}) 
\rangle $, 
$S^f_z(\vec{Q}) = 4 \langle \vec{S}^f_z(\vec{Q}) \cdot  
\vec{S}^f_z(\vec{-Q}) \rangle $, and
$S^f_{xy}(\vec{Q}) = 2 \left( 
           \langle \vec{S}^f_x(\vec{Q}) \cdot \vec{S}^f_x(\vec{-Q}) \rangle  +
	    \langle \vec{S}^f_y(\vec{Q}) \cdot \vec{S}^f_y(\vec{-Q}) \rangle
           \right)$. 
The trial wave function with $ \vec{S}^2 | \Psi_T \rangle \neq 0 $ 
( $ \vec{S}^2 | \Psi_T \rangle = 0 $)
corresponds to the ground state of the Hamiltonian in
Eq.~(\ref{Trial_S0}) (Eq.~(\ref{Trial_N})).  In the {\it large } $\Theta$
limit, the results are independent on the choice of the trial wave function.
In particular, starting from a broken symmetry state  the  symmetry
is restored  at {\it large } values  of $\Theta t $.
For this system, the spin gap is given by $\Delta_{sp} = 0.169 \pm 0.004 $
\cite{Assaad99a}.
Starting with a trial wave function with  $\vec{S}^2 | \Psi_T \rangle \neq 0$,
convergence to the ground state follows approximatively the form:
$ a  + b e^{- \Delta_{sp} 2 \Theta } $. The  solid lines correspond to 
a least square fit to this form. }

\end{figure} 

Fig.~\ref{Conv.fig}  demonstrates the importance of using a spin singlet
trial wave function.   Starting from a N\'eel order for the f-electrons,
convergence to the ground state follows approximatively 
$e^{-\Delta_{sp} 2 \Theta} $ where $\Delta_{sp}$ corresponds to the spin-gap.
When the spin gap is small, convergence is poor   and the remedy is to
consider a spin singlet trial wave function.

\begin{figure}
\begin{center}
\includegraphics[width=9cm,height=6cm]{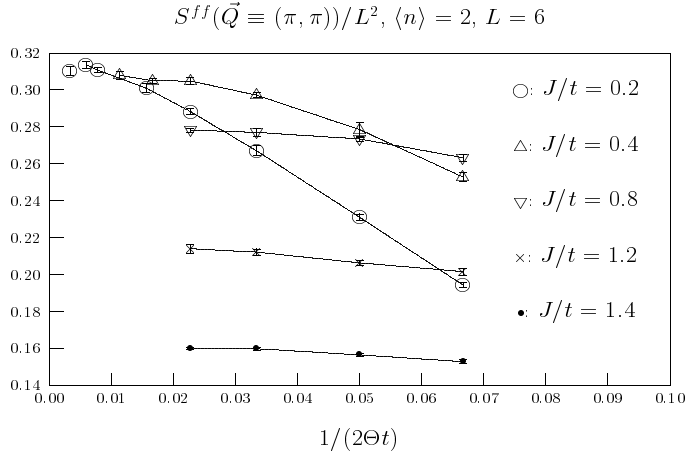}
\end{center}
\caption[]
{\noindent  Spin structure factor   at $\vec{Q} = (\pi,\pi)$ for the $f-$electrons
($S^{ff}(\vec{Q})$)
at various values of $J/t$ and as a function of the projection parameter 
$\Theta t$.  Here we consider a spin singlet trial wave function.
\label{ConvJ.fig}  
}
\end{figure}

Having optimized the trial wave function we now consider convergence as a
function of $J/t$. As apparent from Fig.~\ref{ConvJ.fig}  for small values
of $J/t$ increasingly large  projection parameters are required to obtain 
convergence.  The origin of this  behavior may be traced back to the energy
scale of the RKKY interaction which follows a $J^2$ law. 
At $J/t=0.4$, $2 \Theta t \sim 40$ is enough to obtain convergence whereas at 
$J/t=0.2$,  a value of $2 \Theta t \sim 170 $ is required.

\begin{figure}
\begin{center}
\includegraphics[width=9cm,height=6cm]{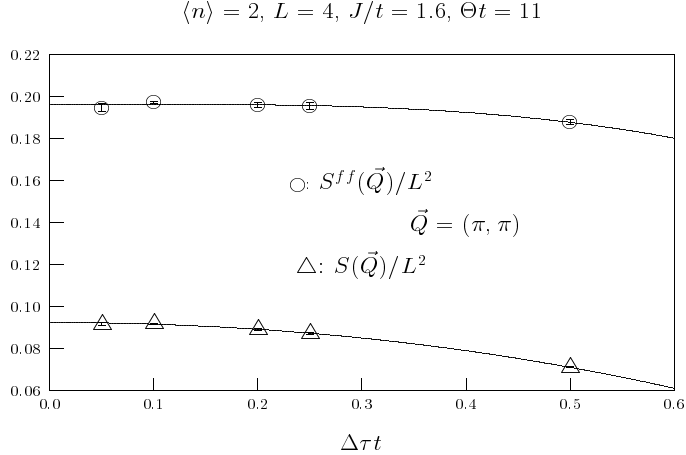}
\end{center}
\caption[]
{\noindent  Systematic error produce by the Trotter decomposition. 
In our simulations, we have used $\Delta \tau = 0.1$ and 
$\Delta \tau = 0.2$. Here, $S(\vec{Q})$ corresponds to the spin structure 
factor of the total spin at $\vec{Q} = (\pi,\pi)$.
\label{Dtau.fig} }
\end{figure}

The systematic error produced by the Trotter decomposition scales as 
$(\Delta \tau )^2$. This behavior is shown in Fig.~(\ref{Dtau.fig}). 
All our calculation were carried out at values of $\Delta \tau$ small
enough so as to neglect this systematic error.

\subsection{Ferromagnetic exchange}

Until now, we have  implicitly considered an antiferromagnetic  exchange,
$J > 0$.
It is straightforward to generalize the above case to a ferromagnetic one. 
The only point to take care of is the choice of the trial wave function in
order  to avoid the sign problem. In this case the non-interacting Hamiltonian 
which generates the trial wave function has to be invariant under the 
particle-hole transformation:
\begin{equation}
\label{PH_F}
	 c^{\dagger}_{\vec{i}, \uparrow} \rightarrow (-1)^{i_x + i_y} 
       c_{\vec{i}, \downarrow}  \; \; {\rm and } \; \; 
         f^{\dagger}_{\vec{i}, \uparrow} \rightarrow +(-1)^{i_x + i_y} 
       f_{\vec{i}, \downarrow}.
\end{equation}
Note that in comparison to Eq.~(\ref{PH}) there is an overall 
sign difference in the 
particle-hole transformation of the f-operators.  With this  condition  one has:
$ \det \left(  U^{<}_{\uparrow,l,s} U^{>}_{\uparrow,l,s} \right)
=   \overline{
    \det \left(  U^{<}_{\downarrow,l,s} 
         U^{>}_{\downarrow,l,s} \right)} $
so that no sign problem occurs.    The trial wave function is thus generated from
the non-interacting Hamiltonian:
\begin{equation}
H_0 =  \sum_{ \vec{k},\sigma }  \varepsilon(\vec{k})
       c^{\dagger}_{\vec{k},\sigma} c_{\vec{k},\sigma}
    - \frac{J}{4} \sum_{\langle \vec{i}, \vec{j} \rangle ,\sigma } 
    (c^{\dagger}_{\vec{i},\sigma} f_{\vec{j},\sigma} +
    f^{\dagger}_{\vec{j},\sigma} c_{\vec{i},\sigma}).
\end{equation}

\subsection{Finite temperature algorithm}
The QMC method presented above may be generalized to finite temperatures
to compute expectation values of observables in the grand-canonical ensemble:
\begin{equation}
  \langle O \rangle = \frac{ {\rm Tr } \left( e^{-\beta H} O \right) }
                           {{\rm Tr}  \left( e^{-\beta H}   \right) }
\end{equation}
Since the step from the $T=0$ approach to the finite-$T$  algorithm is similar
to the one for the standard Hubbard model, we refer the reader to the 
Ref.~\cite{Loh92}.
We note however, that at finite temperatures, the projection onto the $Q_0$
subspace may only be achieved via the inclusion of the Hubbard term
$ U_f \sum_{\vec{i}} ( n^f_{\vec{i},\uparrow} - 1/2 ) 
( n^f_{\vec{i},\downarrow} - 1/2 ) $ in the Hamiltonian. At this point, 
it is very convenient to choose the $SU(2)$-invariant HS decomposition of
Eq.~(\ref{HS2}) since one can take the limit $U_f \rightarrow \infty $
by setting  $\alpha = \pi/2 $. Hence irrespective of the considered 
temperature, we are guaranteed to be in the correct Hilbert space.

\section{Spin and Charge degrees of freedom at $T=0$ }
\label{T0.sec}

\begin{figure}
\begin{center}
\includegraphics[width=12.0cm,height=10.0cm]{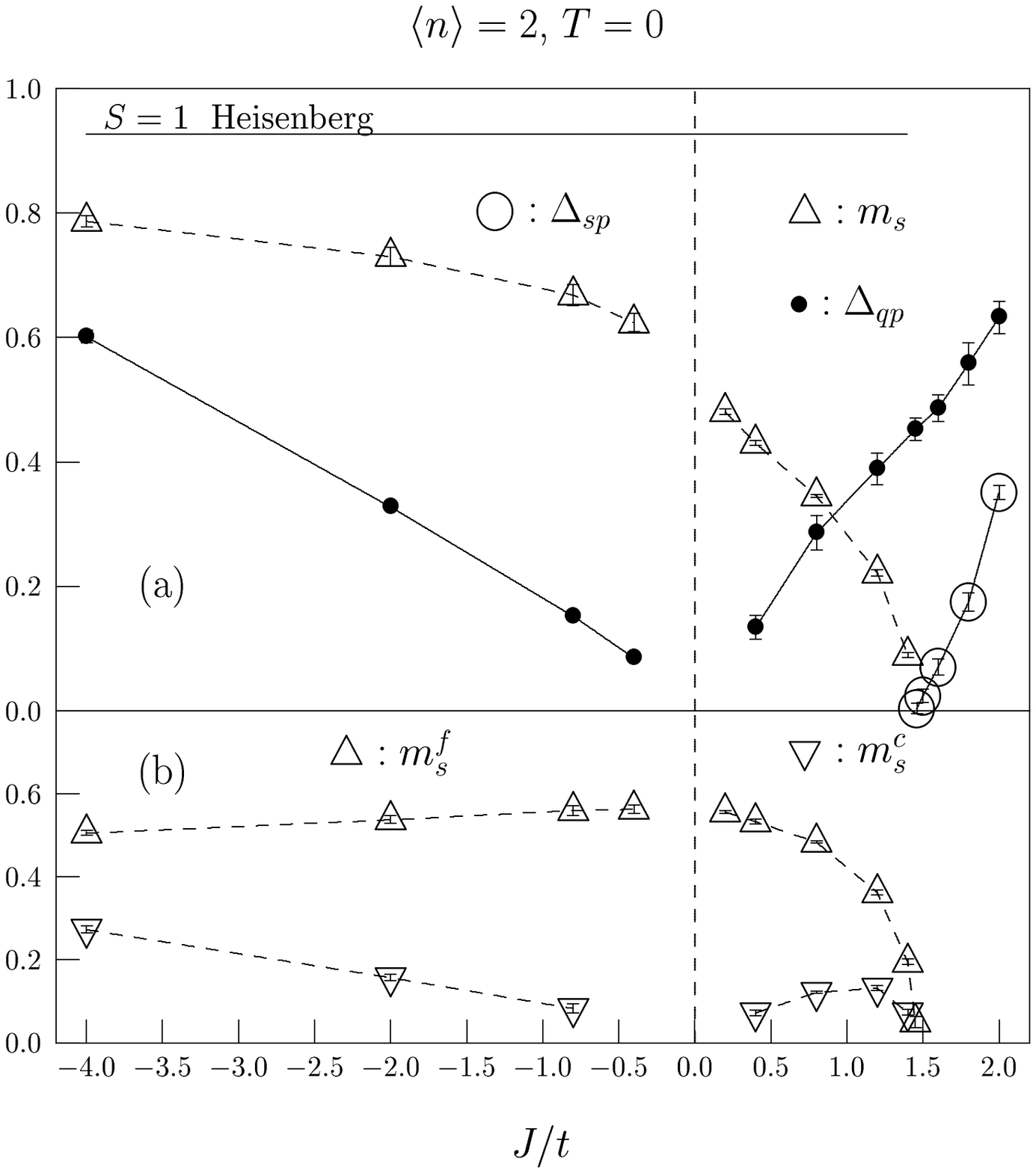}
\end{center}
\caption[]
{\noindent (a) Staggered moment, $m_s$, spin  gap $\Delta_{sp}$ and quasiparticle
gap for the ferromagnetic and antiferromagnetic KLM. All quantities have been
extrapolated to the thermodynamic limit based on results on lattice sizes 
up to $12 \times 12$.
The data for $J>0$ stems from Ref. \cite{Assaad99a}.
The staggered moment corresponds to that of the total spin (see Eq. (\ref{MS_tot})).
The solid line corresponds to the value of the staggered moment for the $s=1$ 
antiferromagnetic model as obtained in a spin wave approximation \cite{Kittel63}.
(b) Staggered moment of the 
$f-$ and $c-$ electrons after extrapolation to the thermodynamic
limit.
\label{Phase.fig}
}
\end{figure}

The different phases  occurring at half-filling are summarized in 
Fig.~\ref{Phase.fig}. All quantities have been extrapolated to the thermodynamic
limit  \cite{Assaad99a}. We have considered sizes ranging from $4 \times 4$ to
$12 \times 12 $ with periodic boundary conditions.
The staggered moment:
\begin{equation}
      m_s = \lim_{L \rightarrow \infty } 
  \sqrt{ \frac{4}{3} \langle \vec{S}(\vec{Q}) \cdot \vec{S}(-\vec{Q})  \rangle }
\label{MS_tot}
\end{equation}
indicates the presence of long-range magnetic order. Here, $\vec{S}(\vec{Q}) 
= \frac{1}{L} \sum_{\vec{j}} e^{i \vec{Q} \cdot \vec{j} } \vec{S}(\vec{j})$ where
$\vec{S}(\vec{j}) = \vec{S}^f(\vec{j}) + \vec{S}^c(\vec{j}) $ is the total
spin, $\vec{Q} = (\pi,\pi)$ the antiferromagnetic wave vector and 
$L$ corresponds to the linear size of the system.  This quantity is maximal at
$J/t = -\infty $ and vanishes at $J_c/t \sim 1.45 $ thus signaling a phase
transition.
The onset of a spin gap,
\begin{equation}
          \Delta_{sp} = \lim_{L \rightarrow \infty} 
                E^L_0(S=1,N_p = 2N) - E^L_0(S=0,N_p = 2N), 
\end{equation}
is observed when magnetic order disappears. Here, $E^L_0(S,N_p )$ is the 
ground state energy on a square lattice with $N=L^2$ unit cells, $N_p$ 
electrons and spin $S$.
Finally, the system remains an insulator for all considered coupling constants.
This is supported  by a non-vanishing quasiparticle gap,
\begin{equation}
          \Delta_{qp} = \lim_{L \rightarrow \infty}
                E^L_0(S=1/2,N_p = 2N + 1) - E^L_0(S=0,N_p = 2N).
\end{equation}
We will first discuss the spin degrees of freedom and then turn our attention 
to charge degrees of freedom. 

\subsection {Spin degrees of freedom}
To investigate  the spin degrees of freedom we compute the dynamical 
spin susceptibility,
\begin{equation}
          S(\vec{q}, \omega) = 
       \pi  \sum_n | \langle n | \vec{S}(\vec{q}) | 0 \rangle |^2 
        \delta ( \omega -(E_n - E_0)).
\end{equation}
where the sum runs over a complete set of eigenstates and $ | 0 \rangle $ 
corresponds to the ground state.  This quantity is related to the imaginary
time spin-spin correlations which we compute with the QMC method
\cite{Assaad99a}:
\begin{equation} 
         \langle 0 | \vec{S}(\vec{q},\tau) \cdot \vec{S}(-\vec{q}) | 0 \rangle 
   = \frac{1}{\pi}\int {\rm d} \omega e^{-\tau \omega } S(\vec{q}, \omega).
\end{equation}
Here, $\vec{S}(\vec{q},\tau) = e^{\tau H } \vec{S}(\vec{q}) e^{- \tau H}$. 
We use the Maximum Entropy (ME) method to accomplish the above  numerically 
ill defined inverse Laplace transform~\cite{Jarrell96}. 

\begin{figure}
\noindent
\mbox{\epsfig{file=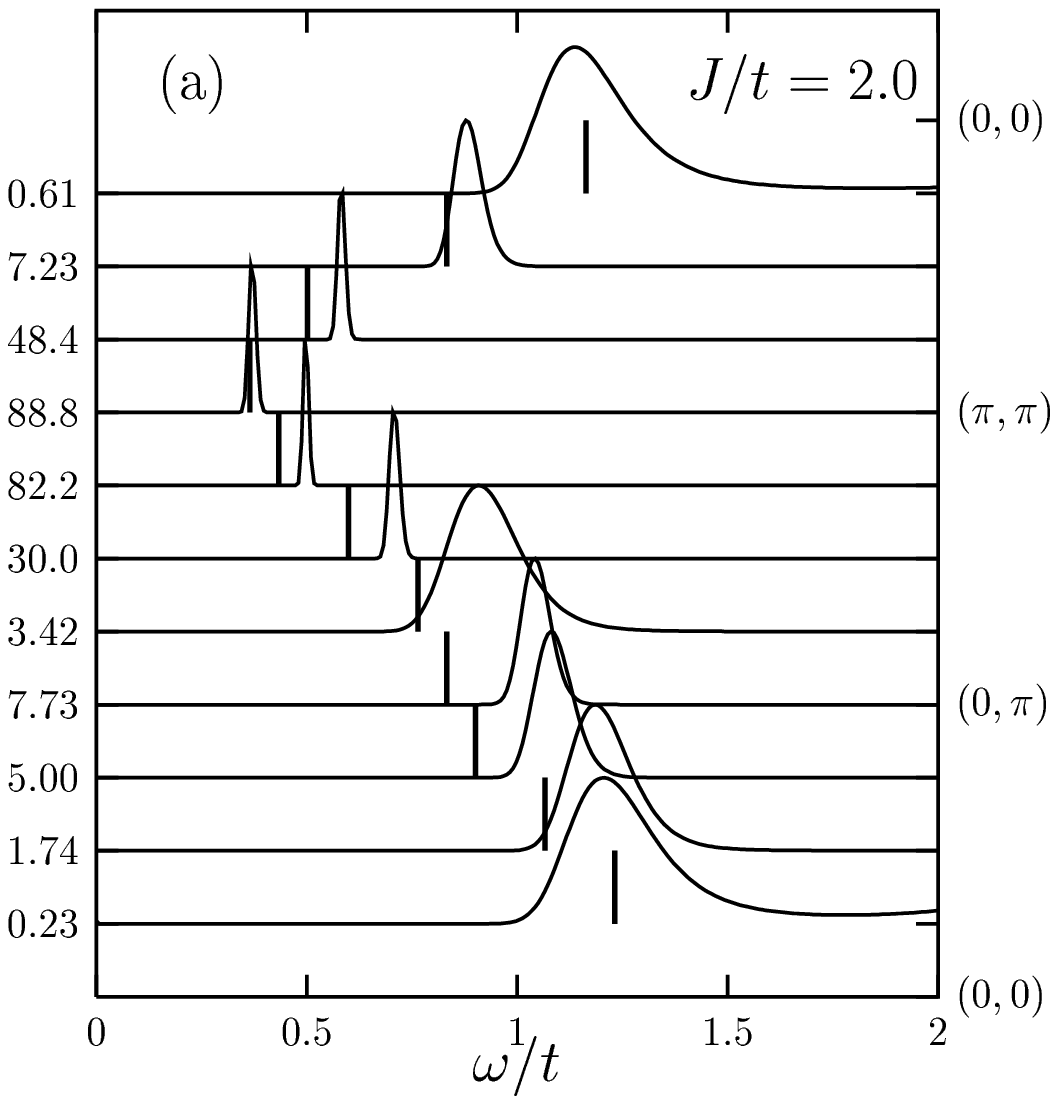, width=7.5cm,height =7.0cm,angle=0} }
\mbox{\epsfig{file=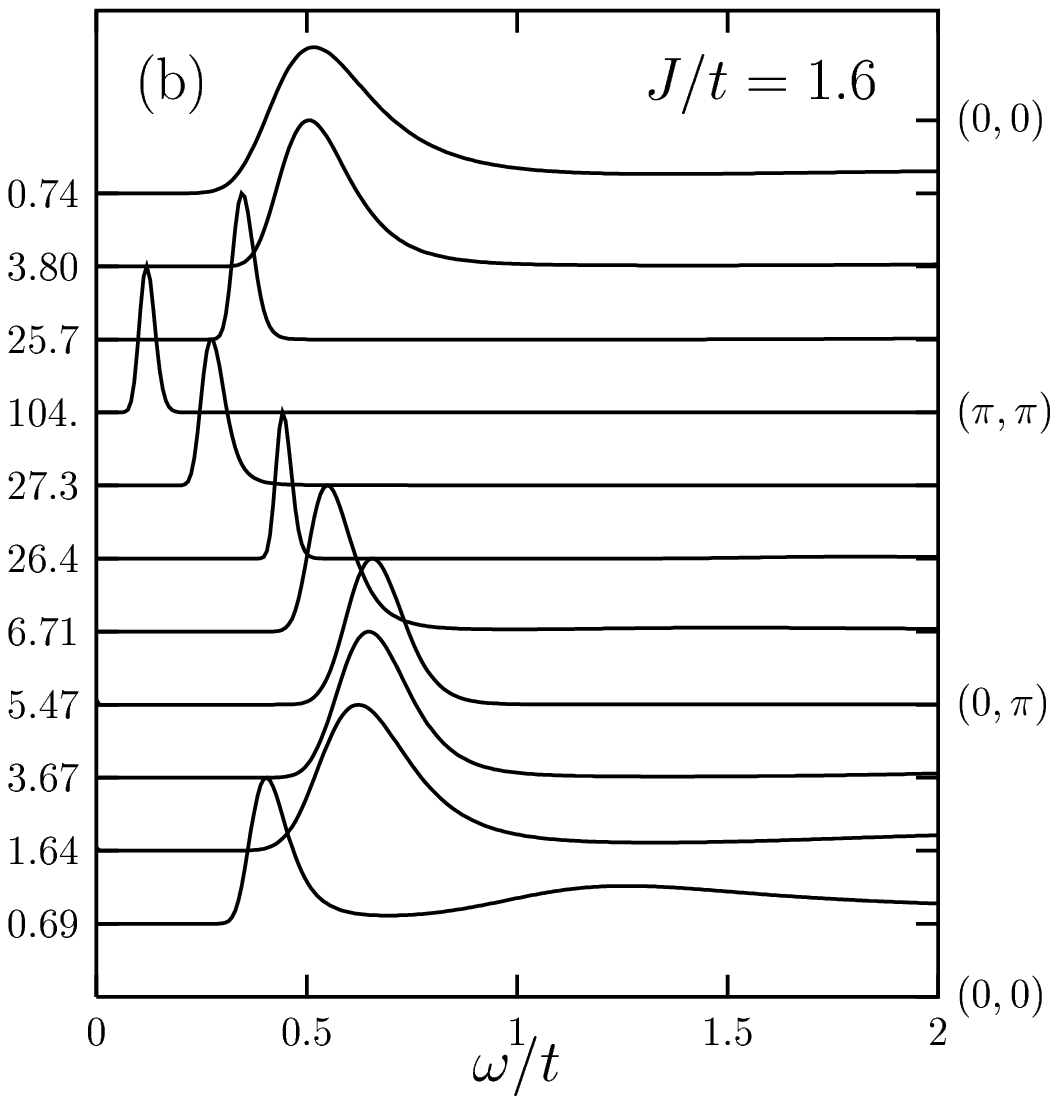, width=7.5cm,height =7.0cm,angle=0} }
\\ \\
\mbox{\epsfig{file=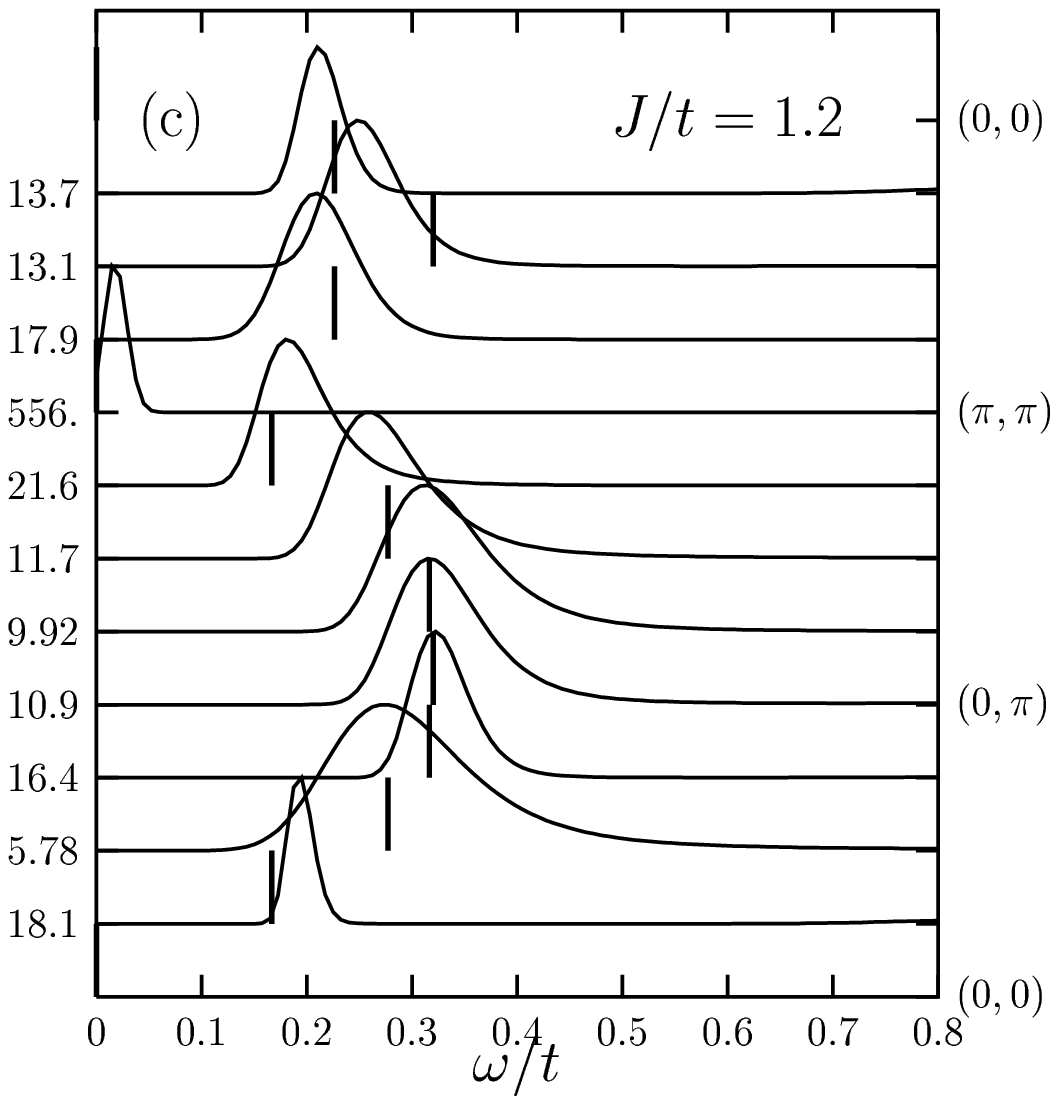, width=7.5cm,height =7.0cm,angle=0} }
\mbox{\epsfig{file=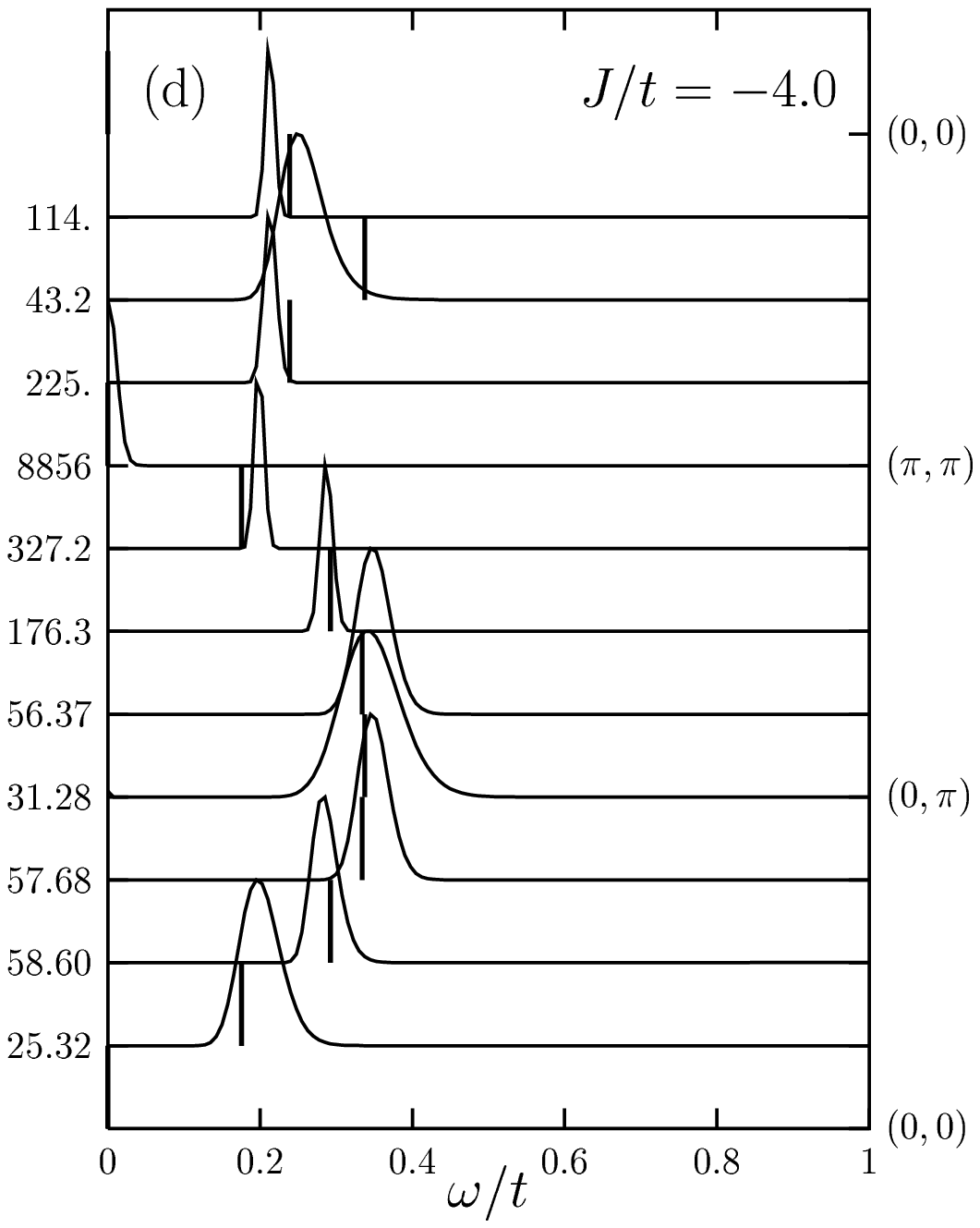, width=7.5cm,height =7.0cm,angle=0} }
\caption[]
{\noindent  Dynamical spin structure factor at $T=0$ for the ferromagnetic and
antiferromagnetic KLM.  We have normalized the peak heights to unity. The numbers
on the left hand side of the figures correspond to the normalization factor. The 
vertical bars are fits to the data, see text.

\label{Spind.fig} }
\end{figure}

In the strong coupling limit $J \rightarrow \infty $, the model becomes trivial,
since each $f-$spin captures a conduction electron to form a singlet.  In 
this limit, the ground state corresponds to a direct product of singlets on
the $f$-$c$ bonds of a unit cell. Starting from this state, one may  create 
a magnon excitation by breaking a singlet to form a triplet. In second-order
perturbation in $t/J$, this magnon acquires a dispersion relation given by: 
\begin{equation}
\label{SC_sp}
        E_{sp}(\vec{q}) = J - \frac{16t^2}{3J} + 
         \frac{4t^2}{J}  \gamma(\vec{q})
\end{equation}
where $\gamma(\vec{q}) = \cos(q_x) + \cos(q_y)$ \cite{Tsunetsugu97_rev}. 
At $\vec{Q} = (\pi,\pi)$,  $E_{sp} (\vec{q}) $ is minimal and is nothing 
but the spin gap.  
In Fig.~\ref{Spind.fig}a, we plot the dynamical spin structure factor for 
$J/t = 2.0$.  The solid bars in the plot correspond to a fit to the above 
strong coupling functional form: $a + b \gamma(\vec{q})$.  As apparent,
this functional form reproduces well the QMC data. We note that this
magnon mode lies below the particle hole continuum  located at 
$2 \Delta_{qp}$ (see Fig.~\ref{Phase.fig}).
 
As we approach the antiferromagnetically ordered phase, one expects that 
the above magnon mode evolves towards a spin-wave form:
\begin{equation}
        E_{sw}(\vec{q}) = \frac{J_{sp}}{2}\sqrt{ 1 - \gamma(\vec{q})^2/4}.
\end{equation}
As apparent from Fig.~\ref{Spind.fig}b as one approaches $J_c$ the spin 
gap  vanishes  and the magnon mode softens around $\vec{q} = \vec{0}$.
In the antiferromagnetic phase (see Fig.~\ref{Spind.fig}c) 
the data follow well the above spin-wave form.

In the limit of large ferromagnetic couplings, the model maps onto the $S=1$ 
antiferromagnetic Heisenberg model.  At $J/t = -\infty$, the ground state
is macroscopically degenerate since the  $f$-$c$ bonds are effectively
decoupled 
and occupied by a triplet with arbitrary  z-component of spin.
This degeneracy is lifted in second-order perturbation theory, yielding
a $S=1$ antiferromagnetic Heisenberg model:
\begin{equation}
        H_{eff} = \frac{2t^2}{J} \sum_{ \langle \vec{i}, \vec{j} \rangle  }
           \vec{\cal{S}}_{\vec{i}} \cdot \vec{\cal{S}}_{\vec{j}}.
\end{equation}
Here,
$ \vec{\cal{S}}_{\vec{i}} =
\sum_{m,m'} t_{\vec{i},m}^{\dagger} \vec{\sigma}^{(1)}_{m,m'} t_{\vec{i},m'} $,
$t_{\vec{i},1} = c^{\dagger}_{\vec{i},\uparrow} f^{\dagger}_{\vec{i},\uparrow}$,
$t_{\vec{i},0} = \frac{1}{\sqrt{2}}
\left( c^{\dagger}_{\vec{i},\uparrow} f^{\dagger}_{\vec{i},\downarrow} +
    c^{\dagger}_{\vec{i},\downarrow} f^{\dagger}_{\vec{i},\uparrow} \right) $, 
and
$t_{\vec{i},-1} = c^{\dagger}_{\vec{i},\downarrow} f^{\dagger}_{\vec{i},
\downarrow} $. 
$\vec{\sigma}^{(1)}$ correspond to the $s=1$ Pauli spin matrices.
The magnetic excitations are clearly spin waves as confirmed by the QMC
data of Fig.~\ref{Spind.fig}d. In the limit of large negative $J$,
the staggered moment should scale to the value obtained for the $S=1$
Heisenberg model. 
Within a spin density wave approximation \cite{Kittel63}, 
this quantity takes the value 0.93.  As apparent
from Fig.~\ref{Phase.fig}, the QMC data approaches smoothly this value as
$J/t$ decreases. 

\begin{figure}
\noindent
\begin{center}
\mbox{\epsfig{file=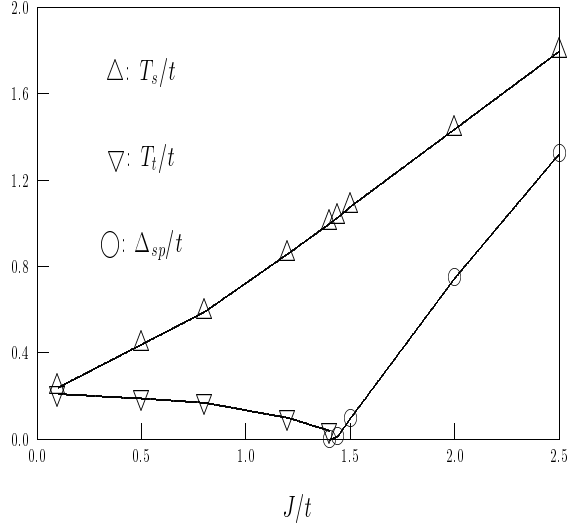, width=7.5cm,height =7.0cm,angle=0} }
\end{center}
\caption[]
{\noindent Mean field   of the two dimensional  Kondo necklace model. 
$T_s$ ($T_t$)  corresponds to the energy scale below which the bond singlets 
(triplets) condense.  $\Delta_{sp}$ denotes the spin gap. 
\label{KN.fig} }
\end{figure}

The quantum phase transition in the spin degrees of freedom at 
$J_c/t \sim 1.45$  may be described in the framework of the Kondo 
necklace model given by:
\begin{equation}
        H_{KN} = t \sum_{\vec{i},\vec{j}}
             \left( S^{c,x}_{\vec{i}} S^{c,x}_{\vec{j}} +  
             S^{c,y}_{\vec{i}}  S^{c,y}_{\vec{j}}\right) +
              J \sum_{\vec{i}} \vec{S}^c_{\vec{i}}  \cdot \vec{S}^f_{\vec{i}}
\end{equation}
This model neglects charge fluctuations, and the spin flip processes 
between conduction electrons mimic the kinetic energy. 
Although the Kondo necklace model has a lower symmetry ($U(1)$) than the KLM
($SU(2)$) one may expect this model
to give a reasonable description of the spin degrees of freedom  at energy scales
smaller than the charge gap.  
A mean-field solution is obtained in terms of bond singlet and triplet
operators~~\cite{Bhatt}. Both the conduction  and impurity spins are 
represented by  singlets,  $\Delta^{\dagger}_{\vec{i}} $, 
and  triplets $ {\vec{t}}^{\phantom{2}\dagger}_{\vec{i}}$ 
on the $f-c$ bonds of the unit-cell. 
The bond operators obey bosonic  commutation rules and are subject to 
the constraint $\Delta^{\dagger}_{\vec{i}} \Delta_{\vec{i}} +
\vec{t}^{\phantom{2}\dagger}_{i} \vec{t}_{i} = 1 $.    
At the mean-field level and generalizing the work of Zhang {\it et
al.}~\cite{Zhang00}  to
finite temperatures, one obtains the phase diagram shown in Fig.~\ref{KN.fig}.
The condensation of singlets $ s = \langle  \Delta^{\dagger}_{\vec{i}} \rangle > 0$
occurs at a temperature scale $T_s$ which, to a first approximation, tracks $J$.
At $J > J_c$ the triplet excitations remain gapped and have a dispersion
relation given by:  
$ \omega(\vec{q}) = \alpha \sqrt{ 1 + s^2 t \gamma(\vec{q})/\alpha } $
with $\alpha = s^2 t \left( 1 + \sqrt{ 1 + \Delta_{sp}^2/t^2 s^2} \right) $.
Here $\Delta_{sp}$ corresponds to the spin gap plotted in Fig.~\ref{KN.fig}.   
The gap in the magnon spectrum at $\vec{q} = (\pi,\pi) \equiv \vec{Q}$ vanishes at 
$J_c/t \sim 1.4 $ in remarkable agreement  
with the QMC results. We note that 
this mean-field approach shows no 
phase transition in the one-dimensional case consistently with numerical calculations~\cite{Tsunetsugu97_rev,Zhang00}. 
For $J < J_c$ the ground state has both 
condensation of singlets ($s > 0$) and of triplets at the antiferromagnetic 
wave vector ( $\bar{t} = \sqrt{N} \langle t^{\dagger,x}_{\vec{Q}} \rangle > 0 $).
The energy scale below which the triplet excitations condense is denoted by
$T_t$ in Fig.~\ref{KN.fig}.  In terms of the KLM, the condensation of 
triplets (singlets)  follows from the RKKY interaction (Kondo effect). Thus,
the fact that at the mean-field level, both $s$ and $\bar{t}$ do not
vanish may be interpreted as coexistence of Kondo screening and 
antiferromagnetism in the ordered phase. We will confirm this point of view
in the study of the charge degrees of freedom.

\subsection{Charge degrees of freedom}
\begin{figure}
\noindent
\mbox{\epsfig{file=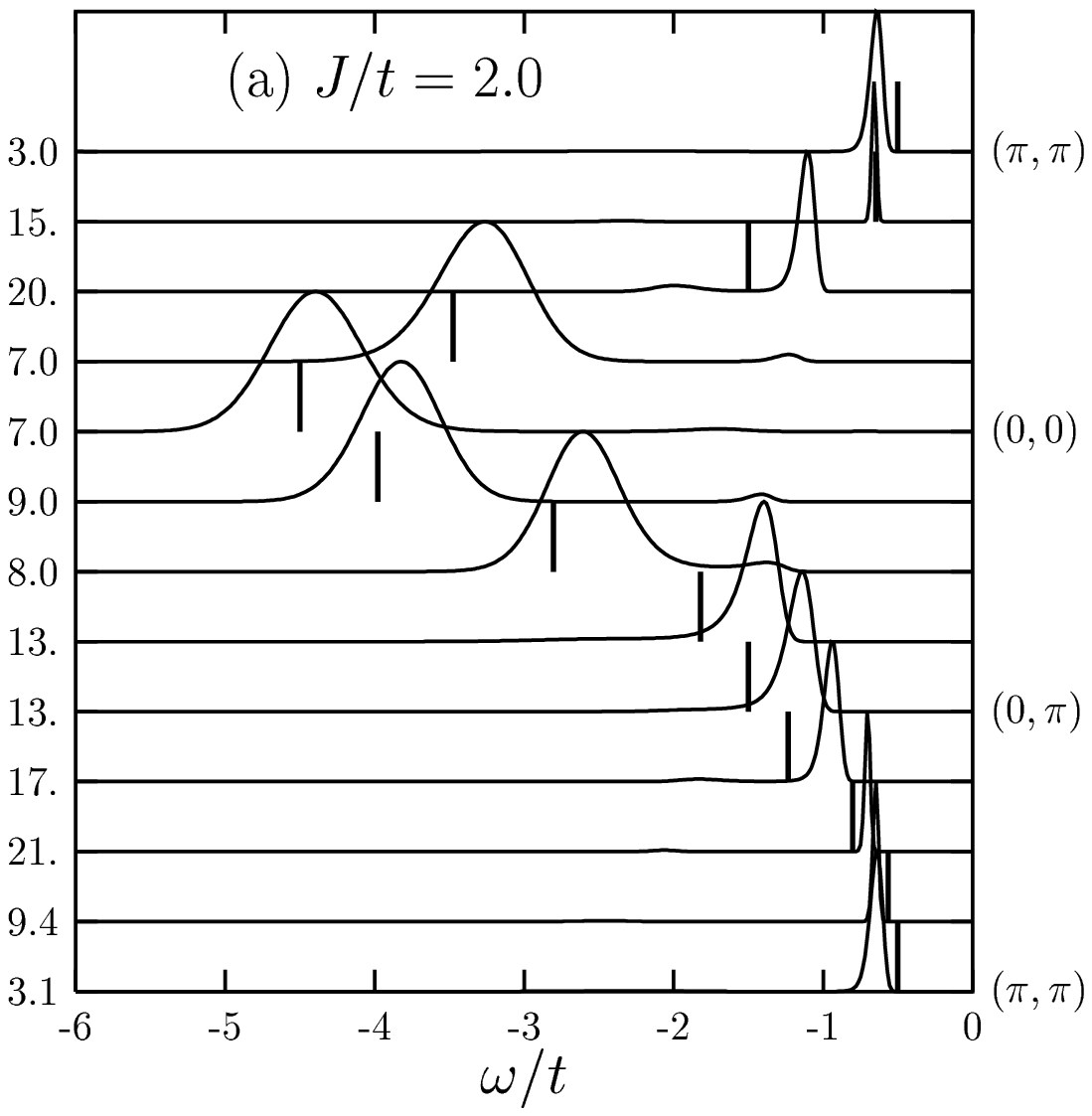, width=7.5cm,height =7.0cm,angle=0} }
\mbox{\epsfig{file=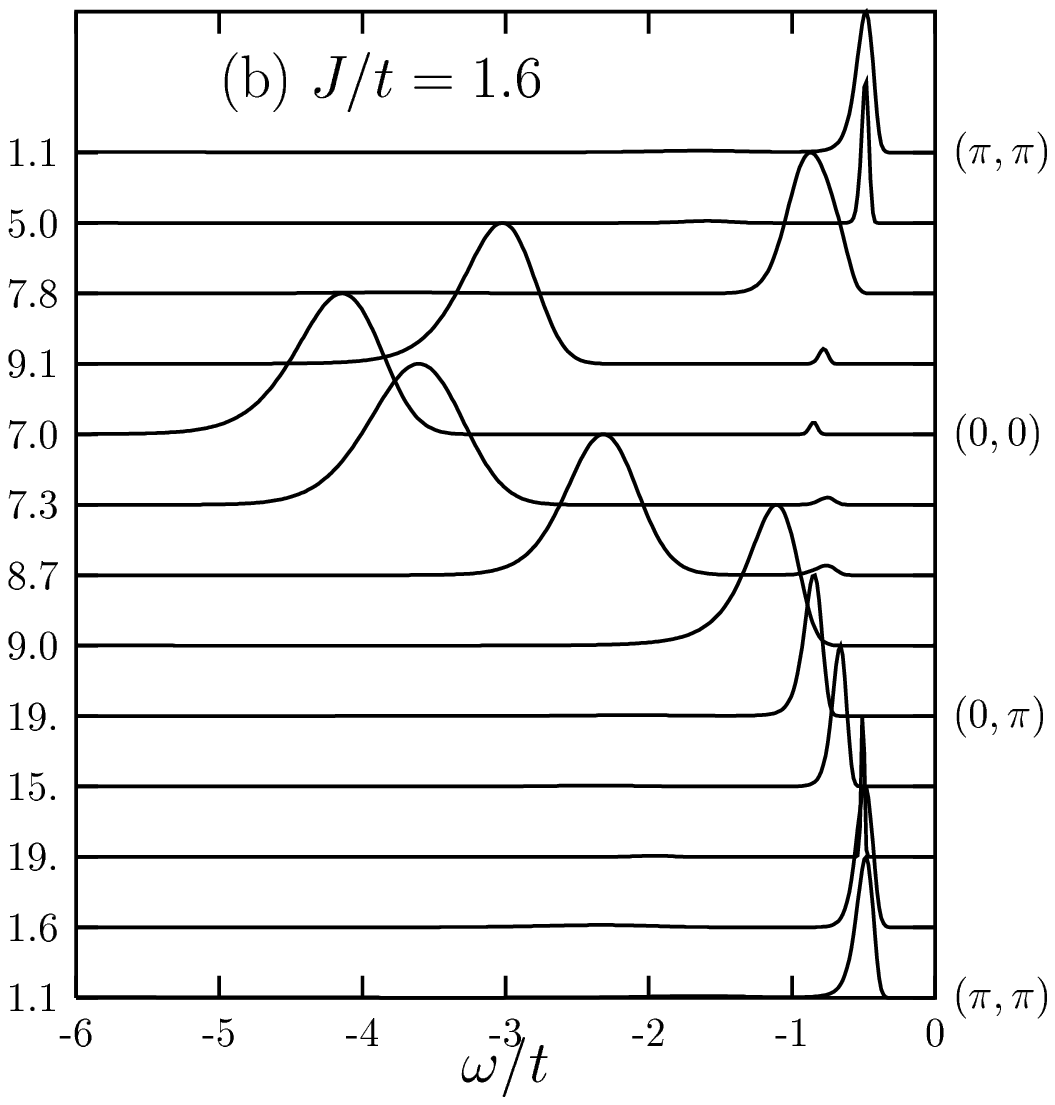, width=7.5cm,height =7.0cm,angle=0} }
\\ \\
\mbox{\epsfig{file=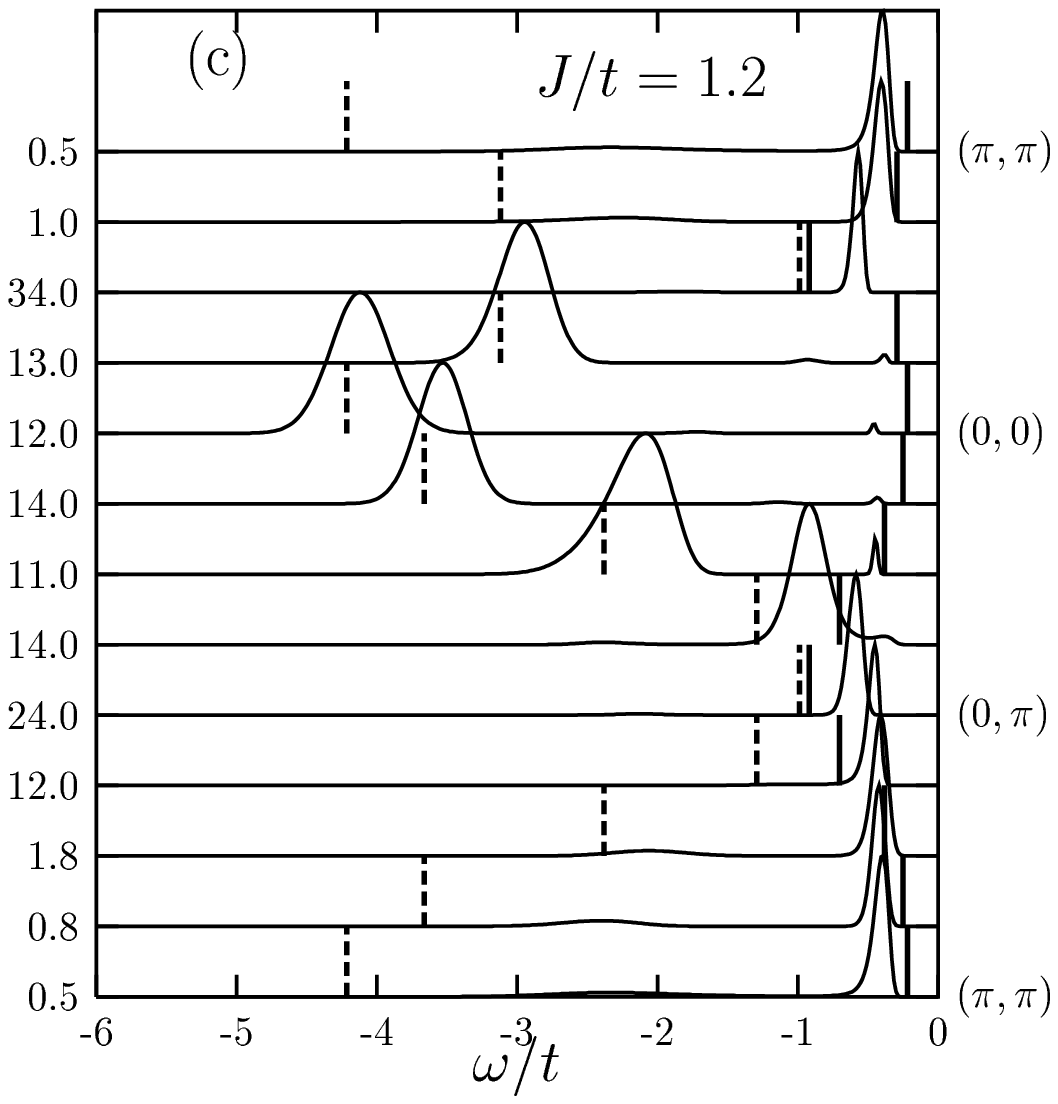, width=7.5cm,height =7.0cm,angle=0} }
\mbox{\epsfig{file=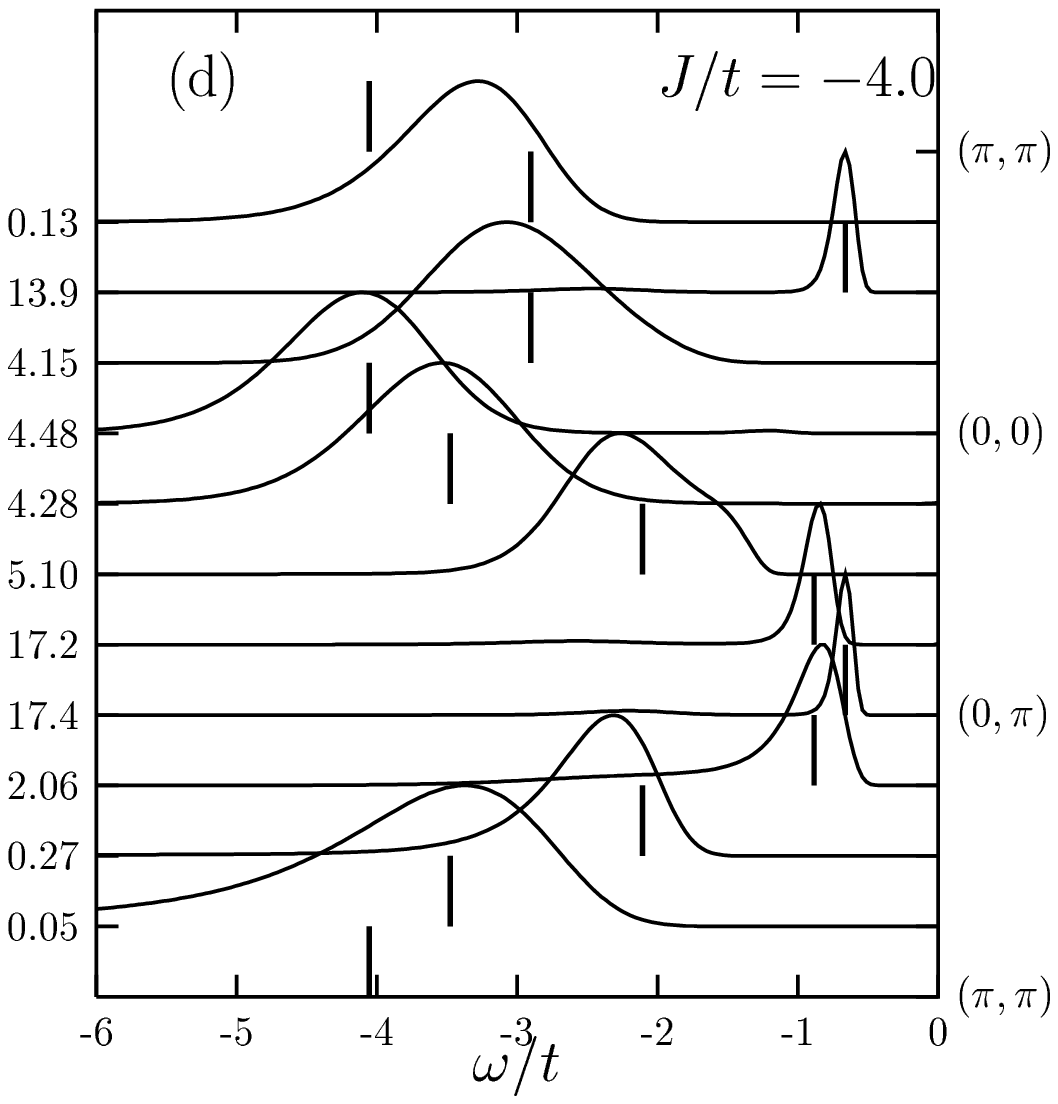, width=7.5cm,height =7.0cm,angle=0} }
\caption[]
{\noindent Single particle spectral function  at $T=0$ for the ferromagnetic and
antiferromagnetic KLM.  We have normalized the peak heights to unity. The numbers
on the left hand side of the figures correspond to the normalization factor. The 
vertical bars are fits to the data, see text.
\label{Akomd.fig} }
\end{figure}

To study the charge degrees of freedom, we compute the spectral function 
$A(\vec{k},\omega)$ which is related to the imaginary time Green function via:
\begin{equation}
       \langle  c^{\dagger}_{\vec{k}}(\tau)     c_{\vec{k}} \rangle
     = \frac{1}{\pi} 
    \int_{0}^{\infty} {\rm d} \omega  e^{-\tau \omega} A(\vec{k}, -\omega).
\label{AKOM}
\end{equation}
The Maximum Entropy (ME) method is used to extract $A(\vec{k},\omega)$.
Starting from the bond-singlet ground state valid in the strong coupling limit, 
one can create a quasiparticle excitation 
which to first order in $t/J$  has the dispersion relation
\begin{equation}
\label{SC_qp}
E_{qp} (\vec{k}) =  3J/4 + t\gamma(\vec{k}).
\end{equation}
$E_{qp} (\vec{k})$  is a minimal at $\vec{k} = (\pi,\pi)$ so that  
the quasiparticle gap
takes the value $\Delta_{qp} =E_{qp} (\vec{k}=(\pi,\pi))$. 
Comparison with Eq.~(\ref{SC_sp}) leads to $\Delta_c = 2 \Delta_{qp} >
\Delta_{sp}$ in the strong coupling limit. This marks the difference to a
standard band insulator which satisfies   $\Delta_c = \Delta_{sp}$. 
In accordance with the strong coupling limit, the numerical data of Fig.~\ref{Akomd.fig} (a)-(c) show that irrespective of 
$J/t$ the quasiparticle gap is defined by the $\vec{k} = (\pi,\pi)$  wave
vector.  Furthermore comparison with Fig.~\ref{Spind.fig} shows that 
the inequality $\Delta_c > \Delta_{sp} $ is valid for all considered coupling
constants. 

\begin{figure}
\begin{center}
\mbox{\epsfig{file=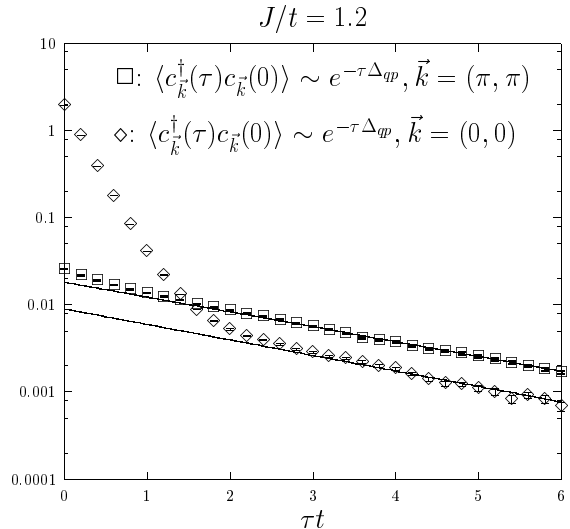, width=7.5cm,height =7.0cm,angle=0} }
\end{center}
\caption[]
{\noindent  $\langle  c^{\dagger}_{\vec{k} + \vec{Q}}(\tau)     c_{\vec{k} +\vec{Q} }
\rangle $ and $\langle  c^{\dagger}_{\vec{k} }(\tau)    c_{\vec{k}  } \rangle $
as a function of $\tau t$ on a $ 8 \times 8 $ lattice at $T=0$.
Both considered $\vec{k}$ points follow a  
$ e^{- \Delta_{qp} \tau } $ law (solid lines) thus confirming the existence of
shadows.  
\label{Shad.fig} }
\end{figure}

For antiferromagnetic couplings ($J > 0$), 
the spectral function shows similar features in the 
ordered ($J < J_c$) and disordered ($J> J_c$) magnetic phases thus lending
support that  Kondo screening which is responsible for the functional
form of the dispersion relation at strong couplings is equally present in the 
ordered phase. However, upon closer analysis, shadow features are seen in 
the antiferromagnetically ordered phase. Shadows are most easily understood 
by considering the self-energy 
$\sum (\vec{k},\omega_m) \propto \frac{g^2}{N \beta} \sum_{\nu_m, \vec{q} }
\chi({\vec{q}},\nu_m)  G_{0}(\vec{k}-\vec{q},\omega_m -
\nu_m) $
describing electrons with propagator $G_{0}(\vec{k}, \omega_m)$ 
scattering off spin fluctuations with coupling constant $g$.
Long-range magnetic order at wave vector $\vec{Q} = (\pi,\pi)$
and staggered moment $m_s$ justify 
the Ansatz $ \chi({\vec{q}},\nu_m) \propto  m^2_s N \beta
\delta_{\nu_m,0} \delta_{\vec{q},\vec{Q}} $ 
for the spin susceptibility. The Green function  is then given by:
$  G(\vec{k},\omega_m)  = 1 / \left( G_0^{-1}(\vec{k},\omega_m)
   - \alpha G_0(\vec{k}+\vec{Q},\omega_m) \right) $ with 
$ \alpha \propto (gm_s)^2 $.  It is then easy to see that 
if  $ G(\vec{k},\omega)$ has a pole at $\omega_0$ then 
$ G(\vec{k} +\vec{Q},\omega) $ also has a pole at $\omega_0$, i.e. the shadow.
Numerically, it is convenient to establish the existence of shadows by
considering the imaginary time Green function.  Fig.~\ref{Shad.fig} plots 
$\langle  c^{\dagger}_{\vec{k}}(\tau)   c_{\vec{k}} \rangle $ for 
$\vec{k} = (\pi,\pi)$. At large values of $\tau t$ this quantity follows an
exponential law $e^{-\tau \Delta_{qp}}$. This exponential decay generates the
pole in $A(\vec{k},\omega)$  at $\omega = -\Delta_{qp}$ (see Eq. (\ref{AKOM}).
As argued above, 
due to the long-range 
antiferromagnetic order one expects  a pole in  $A(\vec{k} + \vec{Q},\omega)$ 
at $\omega =  -\Delta_{qp}$ i.e. the shadow. 
As demonstrated in Fig.~\ref{Shad.fig},  
$\langle  c^{\dagger}_{\vec{k} + \vec{Q}}(\tau) c_{\vec{k} +\vec{Q} } \rangle $  
shows the same asymptotic behavior as 
$\langle  c^{\dagger}_{\vec{k}}(\tau) c_{\vec{k}} \rangle$.  Thus the low energy
feature around $\vec{k} = (0,0) $ in Fig.~\ref{Akomd.fig}c corresponds to 
the shadow of the band  in the vicinity of $\vec{k} = (\pi,\pi) $. We note
that  shadow features at high energies are hard to resolve within the ME.
Close to the phase transition in the disordered phase precursors features of 
the shadow bands are seen (see Fig.~\ref{Akomd.fig}b). As apparent they are
shifted by an energy scale which corresponds approximately to the spin gap.

To obtain further insight into the charge degrees of freedom we will  consider 
a mean-field theory of the KLM recently introduced by Zhang and Yu~\cite{Zhang00b}.
This mean-field theory is appealing since i) it takes into account both 
Kondo screening and magnetic ordering of the $f$  and $c$ electrons
and ii) a phase where both Kondo screening and magnetic ordering emerges 
in a narrow region  around the phase transition. 

Following Zhang and Yu~\cite{Zhang00b} we  write the KLM as:
\begin{eqnarray}
\label{H1_KLM}
 H_{KLM} =  \sum_{\vec{k},\sigma} \varepsilon(\vec{k}) 
  c^{\dagger}_{\vec{k},\sigma} c_{\vec{k},\sigma} + 
  & & \frac{J}{4} \sum_{\vec{i}}
  \left( f^{\dagger}_{\vec{i},\uparrow} f_{\vec{i},\uparrow} - 
         f^{\dagger}_{\vec{i},\downarrow} f_{\vec{i},\downarrow}   \right)
  \left( c^{\dagger}_{\vec{i},\uparrow} c_{\vec{i},\uparrow} - 
         c^{\dagger}_{\vec{i},\downarrow} c_{\vec{i},\downarrow}   \right)  
    + \nonumber \\
 - & & \frac{J}{4} \sum_{\vec{i}} 
\left(   \left(  f^{\dagger}_{\vec{i},\downarrow} c_{\vec{i},\downarrow}
         + c^{\dagger}_{\vec{i},\uparrow} f_{\vec{i},\uparrow}  \right)^2
         + 
        \left(  f^{\dagger}_{\vec{i},\uparrow} c_{\vec{i},\uparrow}
       + c^{\dagger}_{\vec{i},\downarrow} f_{\vec{i},\downarrow}  \right)^2
 \right)
\end{eqnarray}
with the constraint: $f^{\dagger}_{\vec{i},\uparrow} f_{\vec{i},\uparrow} +
f^{\dagger}_{\vec{i},\downarrow} f_{\vec{i},\downarrow} = 1 $.   The 
second term of Eq.~(\ref{H1_KLM}) describes the polarization of the conduction electrons by the impurity 
spins and leads to a magnetic instability. The third term 
term is nothing but a rewriting of the spin-flip processes:
\begin{equation}
       \frac{J}{2} \sum_{\vec{i}} 
       \left( f^{\dagger}_{\vec{i},\uparrow} f_{\vec{i},\downarrow} 
              c^{\dagger}_{\vec{i},\downarrow} c_{\vec{i},\uparrow}  +
              f^{\dagger}_{\vec{i},\downarrow} f_{\vec{i},\uparrow} 
              c^{\dagger}_{\vec{i},\uparrow} c_{\vec{i},\downarrow} \right), 
\end{equation}
which are at the 
origin of the screening of the impurity spins by the conduction electrons. 
The mean-field approximation proposed by  Zhang  and Yu~\cite{Zhang00b} is based on the 
order parameters:
\begin{eqnarray}
\label{MFoder}
  & &  \langle f^{\dagger}_{\vec{i},\uparrow} f_{\vec{i},\uparrow} -
                  f^{\dagger}_{\vec{i},\downarrow} f_{\vec{i},\downarrow}  
           \rangle = 
       m_f e^{i \vec{Q} \cdot \vec{i} } \nonumber \\
  & & \langle c^{\dagger}_{\vec{i},\uparrow} c_{\vec{i},\uparrow} -
       c^{\dagger}_{\vec{i},\downarrow} c_{\vec{i},\downarrow}  \rangle =
      - m_c e^{i \vec{Q} \cdot \vec{i} } \; \; \; {\rm and} \nonumber \\
  & & \langle f^{\dagger}_{\vec{i},\downarrow} c_{\vec{i},\downarrow}
              + c^{\dagger}_{\vec{i},\uparrow} f_{\vec{i},\uparrow} \rangle 
     = \langle  f^{\dagger}_{\vec{i},\uparrow} c_{\vec{i},\uparrow}
              + c^{\dagger}_{\vec{i},\downarrow} f_{\vec{i},\downarrow} 
       \rangle = -V.
\end{eqnarray}
Here $\vec{Q}$ is the antiferromagnetic wave vector, $m_f$ and $m_c$ are  respectively 
the staggered 
moments of the 
impurity spins and conduction electrons and $V$ is the 
hybridization order 
parameter which leads to the screening of the impurity spins. 
With the above Ansatz one obtains the mean field Hamiltonian:
\begin{equation}
\tilde{H} = \sum_{\vec{k},\sigma} \left( \begin{array}{c}
c_{\vec{k},\sigma} \\
c_{\vec{k} + \vec{Q},\sigma} \\
f_{\vec{k},\sigma} \\
f_{\vec{k} + \vec{Q},\sigma}
\end{array} \right)^{\dagger}
\left( \begin{array}{cccc}
\varepsilon(\vec{k})    & \frac{J m_f  \sigma}{4}  & \frac{JV}{2} & 0            \\
\frac{J m_f  \sigma}{4} & -\varepsilon(\vec{k})    & 0            & \frac{JV}{2}  \\
\frac{JV}{2}            & 0                      & 0   & -\frac{J m_c  \sigma}{4} \\
0 & \frac{JV}{2} & -\frac{J m_c  \sigma}{4} & 0
\end{array} \right)
\left( \begin{array}{c}
c_{\vec{k},\sigma} \\
c_{\vec{k} + \vec{Q},\sigma} \\
f_{\vec{k},\sigma} \\
f_{\vec{k} + \vec{Q},\sigma}
\end{array} \right)   + NJ \left( m_f m_c/4 + V^2/2 \right)
\end{equation}
where the $\vec{k}$ sum runs over the magnetic Brillouin zone. We note that due 
to particle-hole symmetry present in the half-filled case, the constraint
of no double occupancy of the $f$-sites is satisfied on average:
$ \langle f^{\dagger}_{\vec{i},\uparrow} f_{\vec{i},\uparrow} +
f^{\dagger}_{\vec{i},\downarrow} f_{\vec{i},\downarrow} \rangle  = 1 $.  
The saddle  point 
equations, 
\begin{equation}
\label{MF_Zhang}
 \langle \frac{\partial \tilde{H}}{\partial m_f} \rangle =
\langle \frac{\partial \tilde{H}}{\partial m_c} \rangle = 
\langle \frac{\partial \tilde{H}}{\partial V} \rangle = 0
\end{equation}
may then be solved \cite{Zhang00b}. 
\begin{figure}
\begin{center}
\mbox{\epsfig{file=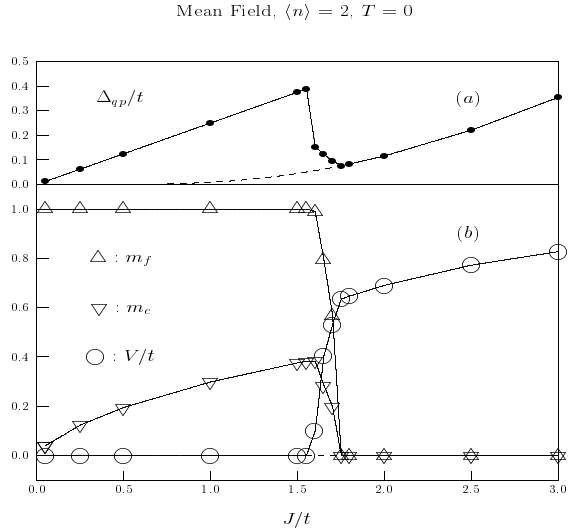, width=7.5cm,height =7.0cm,angle=0} }
\end{center}
\caption[]
{\noindent  Solution of the  mean-field equations in  Eq. (\ref{MF_Zhang}). 
The solid line in (a) corresponds to the quasiparticle gap  as obtained with the mean field 
order parameters plotted in (b).  The dashed line corresponds to  the quasiparticle gap 
obtained in the absence of magnetic ordering (i.e. we consider the solution of the 
mean-field equations with $m_c, m_f = 0$ but $V \neq 0$. In the weak coupling limit, those
solutions produce higher energy values than when magnetic ordering is allowed.)
\label{mean_yu.fig} }
\end{figure}
Solutions to the saddle point equations at $T = 0$ and as a function 
of $J/t$ are plotted in Fig.~\ref{mean_yu.fig}. 
As apparent, solutions with $m_c, m_f \neq 0$, $V=0$  
as well as with  $ m_c, m_f = 0$, $V \neq 0 $ and most 
interestingly with $m_c, m_f, V \neq 0$ are obtained.
Each solution  predicts very different functional forms for the quasiparticle
dispersion relation of the conduction electrons. 
Thus by comparing with the numerical data, we can deduce 
which values of the mean-field order parameters are appropriate to best
describe each phase. 

We start by considering the spin-gap phase with $J > J_c$. Here, magnetic
order is absent  and the  impurity  spins are completely screened by the conduction
electrons. It is thus appropriate to set $  m_c = m_f = 0$ but $ V \neq 0$. This
yields two  quasiparticle bands with dispersion relation:
\begin{equation}
    E_{\pm}(\vec{k})=\frac{1}{2} \left( \varepsilon(\vec{k}) \pm E(\vec{k}) \right), \; \;
{ \rm with} \; \;  E(\vec{k}) = \sqrt{ \varepsilon(\vec{k})^2 + (JV)^2 }.
\end{equation}
The quasiparticle weights are given by the coherence factors:
$ u_{\pm}(\vec{k})^2 = \frac{1}{2}
\left( 1 \pm  \frac{ \varepsilon(\vec{k})}{E(\vec{k})} \right) $.
We can use this form to fit the QMC data shown in Fig.~\ref{Akomd.fig} (a). 
As apparent, the functional form of the dispersion relation is well reproduced.

We now consider  $J < J_c$. Here, antiferromagnetic order is present both in the 
conduction electrons and localized  spins so that: $ m_c \neq 0$ as
well as $m_f  \neq 0$. Following the idea that the spin degrees of freedom
are frozen due to the magnetic ordering ordering, we set $V = 0$ to obtain:
\begin{equation}
\label{E_SDW}
E_{\pm}(\vec{k})= \pm E(\vec{k}) \; \; {\rm with } \; \; 
E(\vec{k}) = \sqrt{ \varepsilon(\vec{k})^2 + (Jm_f/4)^2 } 
\end{equation}
The residues of the poles of the Green function follow: 
$ u_{\pm}(\vec{k})^2 = \frac{1}{2}
      \left( 1 \pm  \frac{ \varepsilon(\vec{k})}{E(\vec{k})} \right). $
This clearly does not reproduce the QMC results since the 
very flat quasiparticle band observed numerically around  $\vec{k}=(\pi,\pi)$ 
is absent (see Fig.~\ref{Akomd.fig}(c)). 
Assuming on the other hand that magnetic ordering and Kondo screening
coexist, we set $V \neq 0$ 
to  obtain four quasiparticle bands:
\begin{eqnarray}
\label{Epmpm}
       E_{\pm,\pm}(\vec{k}) & = &\pm \frac{1}{\sqrt{2} }
\left\{ E(\vec{k}) \pm \sqrt{ E(\vec{k})^2 
- \frac{J^4}{4} (m_cm_f/4 + V^2)^2 - J^2 m_c^2/4 \varepsilon(\vec{k})^2 }     \right\}^{1/2}
{\rm with} \nonumber \\
E(\vec{k}) & = &
\varepsilon(\vec{k})^2   + J^2(m_c^2/4 + m_f^2/4 + 2 V^2)/4
\end{eqnarray}
An acceptable account of the numerical data is obtained by using the 
QMC values of the 
staggered moments and $V$ as a fit parameter  (see Fig.~\ref{Akomd.fig}(c)).
We are thus led to the interpretation that the localized spins play a dual role. 
On one hand they are partially screened
by the conduction electrons. On the other hand the remnant magnetic moment orders 
due to the RKKY interaction. 

It is now interesting to consider the ferromagnetic KLM. When $J<0$, Kondo screening
is not present. Thus, we expect the appropriate mean-field solution to have
$ m_c \neq 0$ as well as  $m_f  \neq 0 $ but $ V = 0$.
This choice of mean-field parameters leads to the dispersion relation given in
Eq.~(\ref{E_SDW}).  As apparent, and using $m_f$ as a fit parameter, we can  
reproduce the QMC results (see Fig.~\ref{Akomd.fig}d).

\section{Spin and Charge degrees of freedom at finite temperature}
\label{FT.sec}

The aim of this section is to define relevant energy scales for both spin 
and charge degrees of freedom as a function of $J/t$.
In doing so, we will discuss the behavior of the 
optical conductivity, staggered spin susceptibility, single particle 
spectral functions as well as 
specific heat  
as a function of temperature. We  will put the emphasis on the behavior of 
those quantities at the  spin and charge energy scales.

\subsection{Spin and Charge energy scales.}
\begin{figure}
\psfrag{TJ}{$T/J$}
\psfrag{Jchi}{$J \chi_c$}
\begin{center}
\includegraphics[width=10cm]{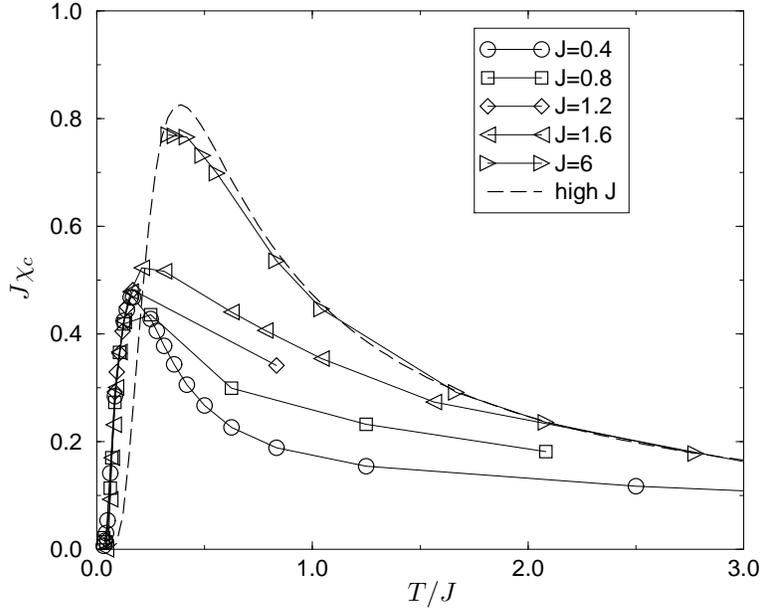}
\end{center}
\caption{Charge susceptibility $J \chi_c$ vs $T/J$ for various couplings on
the $L=6$ lattice. For very large $J$, we obtain good agreement with the large-$J$
expression~(\ref{chic_largeJ.eq}) plotted with a dashed line. }
\label{chic.fig}
\end{figure}

To define the charge scale, we consider the  charge susceptibility 
$\chi_c=\frac{\beta }{ L^2 } \left(\langle N^2\rangle -
\langle N\rangle^2 \right) $  
where $N$ corresponds to the particle number operator.
It suffices to consider only the conduction electrons 
since  the $f$-electrons are localized and have no charge fluctuations.
Since we are discussing  the  temperature dependence of $\chi_c$, 
let us recall the high-temperature result:
\begin{equation}
\chi_c=\frac{1}{2T}\left(1-\frac{1}{8T^2}\left(\frac{3J^2}{8}+8t^2\right)\right).
\label{chic_hiT.eq}  
\end{equation}
From that behavior, it appears that $J\chi_c$ will exhibit some approximative
scaling form as a function of $T/J$ only for large $J/t\gg 8/\sqrt{3}\sim 4.62$. In
Fig.~\ref{chic.fig}, we find consistent results at high temperature with
 (\ref{chic_hiT.eq}).

\begin{figure}
\begin{center}
\includegraphics[width=10cm]{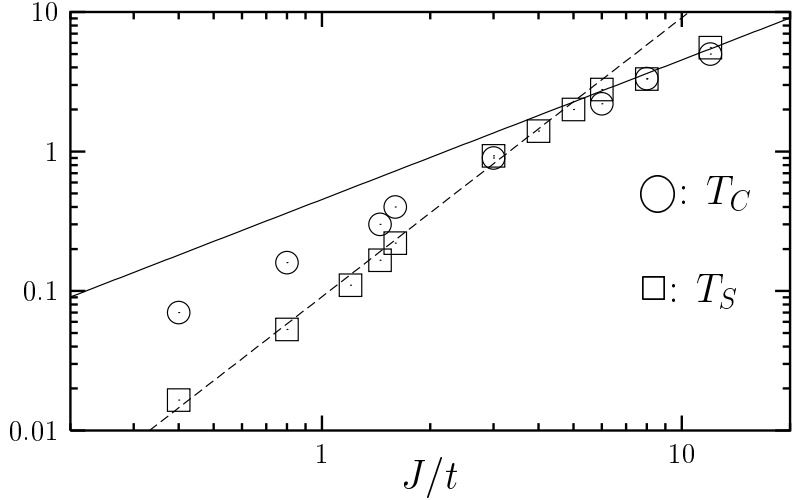}
\end{center}
\caption{Characteristic spin $T_{S}$ and charge $T_{C}$ temperatures as
defined by the maximum of
$\chi_s$ and $\chi_c$  as a function of $J$
for  $L=6$ and $L=8$ when available. At
large $J$, the asymptotic behavior of $T_S$ is $0.453 J$
(Eq.~(\ref{chis_largeJ.eq}) shown in full line)  
with no adjustable parameter; at small $J$, $T_{S}$ is well fitted 
by the form  $J^2/11$ (dashed line).}
\label{Tmag.fig}
\end{figure}

We can define a characteristic charge temperature, $T_C$, in a precise way by 
looking 
at the maximum of $\chi_c$. In the weak coupling limit, our numerical results 
are consistent with $T_C \sim J$ (See Figs.~\ref{chic.fig} and \ref{Tmag.fig}).
In the large $J$ limit ($J>$ bandwidth), the physics becomes local and one can 
consider 
decoupled sites. For each site, there are only 8 states to take into account 
for computing the grand-canonical partition function: the singlet state, the 
three-fold degenerate triplet, the four-fold degenerate $S=1/2$ state containing 
either an empty conduction site or a doubly-occupied one and with the two different 
spin configurations. In this limit, $\Delta_{sp}=J$, $\Delta_{qp}/J=3/4$ and 
\begin{equation}
\chi_c=\beta\frac{4}{4+3e^{-\beta J/4}+e^{3\beta J/4}}
\label{chic_largeJ.eq}
\end{equation}
which exhibits a peak at $T_C=0.386 J$.  
Hence, and apart from different numerical  prefactors at weak and 
strong couplings,
$T_C$ scales as $J$ in both limits (see Fig.~\ref{Tmag.fig}).

To best understand the meaning of the charge scale, we consider the
real part of the  optical 
conductivity as obtained from the Kubo formula, $\sigma(\omega,T)$. This quantity
is related to the imaginary time current-current correlation functions via:
\begin{equation}
\langle J(\tau)J(0) \rangle = \int d \omega K(\omega,\tau) \sigma(\omega,T)
\; \; {\rm with} \; \; K(\omega,\tau) = \frac{1}{\pi}
\frac{e^{-\tau \omega} \omega}{1 - e^{- \beta \omega} }.
\label{kubo.eq}
\end{equation}
Here $J$ is the current operator along  the $x$ or $y$ lattice direction and 
$\langle \rangle$ represents an
average over the  finite-temperature ensemble. The above inverse Laplace transform,
to obtain the optical conductivity is carried out with the ME \cite{Jarrell96} method.
The default model is chosen as follows. 
We start at high temperature with a flat default and then, for lower temperatures, 
we take as default the result obtained at the temperature just 
above~\cite{Gunnarsson00}.
This allows us to obtain smoother results but  emphasizes the fact that the 
ME method depends on the default which is used. 

\begin{figure}
\psfrag{x}{\huge $\omega/t$}
\psfrag{beta2}{\huge $\beta t=2$}
\psfrag{beta5}{\huge $\beta t=5$}
\psfrag{beta12}{\huge $\beta t=12$}
\psfrag{beta18}{\huge $\beta t=18$}
\psfrag{beta40}{\huge $\beta t=40$}
\psfrag{y2}{\huge $2.34$}
\psfrag{y5}{\huge $3.24$}
\psfrag{y12}{\huge $2.95$}
\psfrag{y18}{\huge $3.52$}
\psfrag{y40}{\huge $4.19$}
\begin{center}
\includegraphics[angle=0,width=10cm]{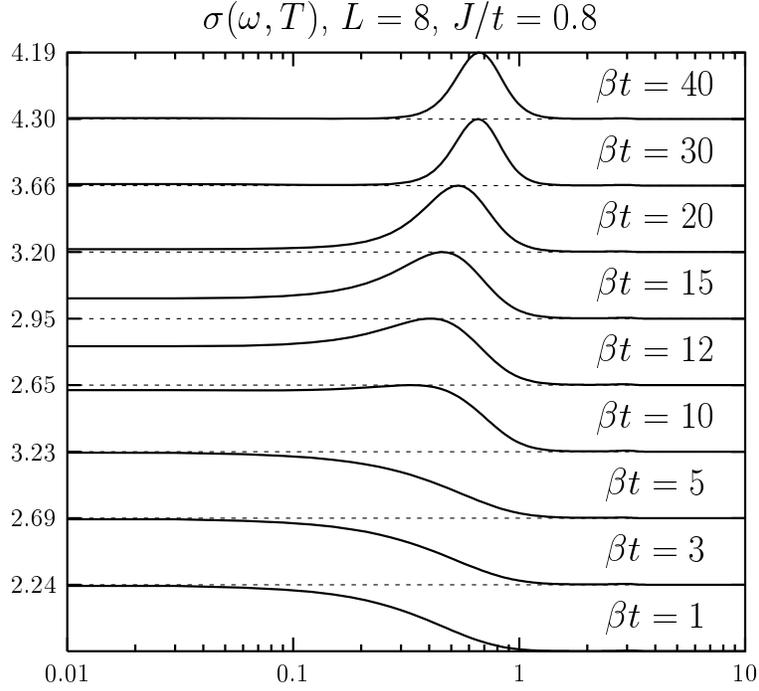}
\end{center}
\caption{Optical conductivity $\sigma(\omega,T)$ versus $\omega$ on
a logarithmic scale for
$J/t=0.8$ and various temperatures ($\Delta\tau=0.2$, $L=8$ lattice). The 
peak height has been normalized to unity and the normalization factor
is listed on the left hand side of the figure. 
As the temperature is decreased below the charge scale, $T_C/t \sim 0.16$, 
spectral weight
is transferred from the Drude peak to finite frequencies.}
\label{sigma.fig}
\end{figure}

The overall features of the conductivity are shown in Fig.~\ref{sigma.fig} for a
given $J$. At high temperatures, there is only a very broad lorentzian Drude peak.
By lowering the temperature, we first observe an enhancement of the
Drude weight as expected for a metal. At temperatures scales lower than
$T_{C}$, there is a transfer of spectral weight from the Drude peak to finite
frequencies and finally, at very low temperatures,  we observe the opening
of an optical gap related to the quasi-particle gap observed in the single 
particle density of states (DOS).  

\begin{figure}
\psfrag{ylabel}{$\rho/\rho_{min}$}
\psfrag{TJ}{\hspace{1cm} $T/J$}
\begin{center}
\includegraphics[width=10cm]{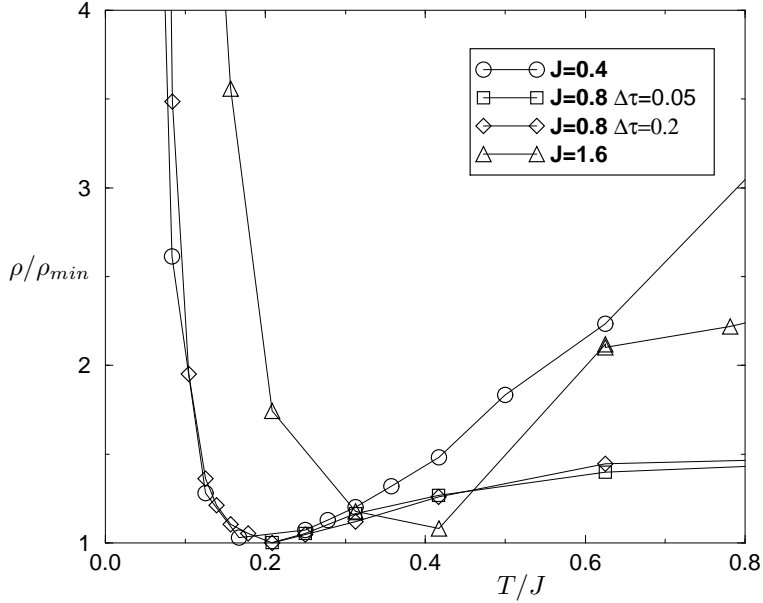}
\end{center}
\caption{Normalized dc resistivity $\rho$ as a function of $T/J$ for various
couplings. We have checked for $J/t=0.8$ that the results
do not depend on $\Delta \tau$. To a first approximation, and 
taking into account the scatter of the data at $J/t = 1.6$,  the temperature
of the minimum in $\rho$ 
tracks $T_C$: $T_C/J \sim 0.2, 0.2, 0.25 $ for $J/t = 0.4, 0.8$
and $1.6$ respectively. 
}
\label{rho.fig}
\end{figure}

The resistivity is defined as $\rho(T)=1/\sigma(0,T)$. In Fig.~\ref{rho.fig}, we
plot $\rho(T)$ for various $J$. We observe a  minimum located
at approximately $T_C$. 
Thus, we will conclude that $T_C$ corresponds to an energy scale where scattering
of the electrons is enhanced while decreasing temperature due
to the screening of magnetic impurities. 

\begin{figure}
\psfrag{JT02xxxx}{$J/t=0.2$}
\psfrag{JT04}{$J/t=0.4$}
\psfrag{JT08}{$J/t=0.8$}
\psfrag{JT16}{$J/t=1.6$}
\psfrag{ylabel}{$C$}
\psfrag{xlabel}{$T/J$}
\begin{center}
\includegraphics[width=10cm]{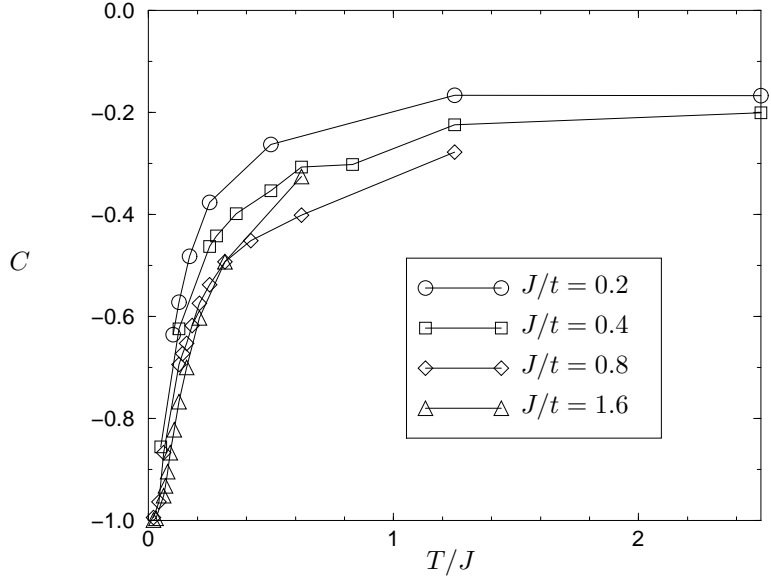}
\end{center}
\caption{Normalized local spin-spin correlation function $C=\langle \vec{S}_f \cdot
\vec{S}_c \rangle(T) / |\langle \vec{S}_f \cdot \vec{S}_c \rangle(T=0)|$ 
as  a function of $T/J$ for various
couplings. To a first approximation, the temperature scale 
of the onset of correlations tracks $J$ . }
\label{SfSc.fig}
\end{figure}

This scenario is reinforced by the behavior of the local spin-spin
correlation function  $C=\langle \vec{S}_f \cdot
\vec{S}_c \rangle(T) / |\langle \vec{S}_f \cdot \vec{S}_c \rangle(T=0)|$
plotted in Fig.~\ref{SfSc.fig}. As the temperature is lowered, this
quantity decreases indicating the formation of local singlets. Since the
curves are almost identical as a function of $T/J$ and for various
couplings, we deduce that the typical energy scale is $J$ and that
the formation of those singlets are responsible for the enhancement of the 
resistivity which occurs at a  similar temperature.

\begin{figure}
\psfrag{Ek}{$E(k)$}
\psfrag{k}{$k$}
\psfrag{a}{$\Delta_{qp}$}
\psfrag{b}{$\Delta_{opt}$}
\psfrag{pi}{$\pi$}
\begin{center}
\includegraphics[angle=-90,width=10cm]{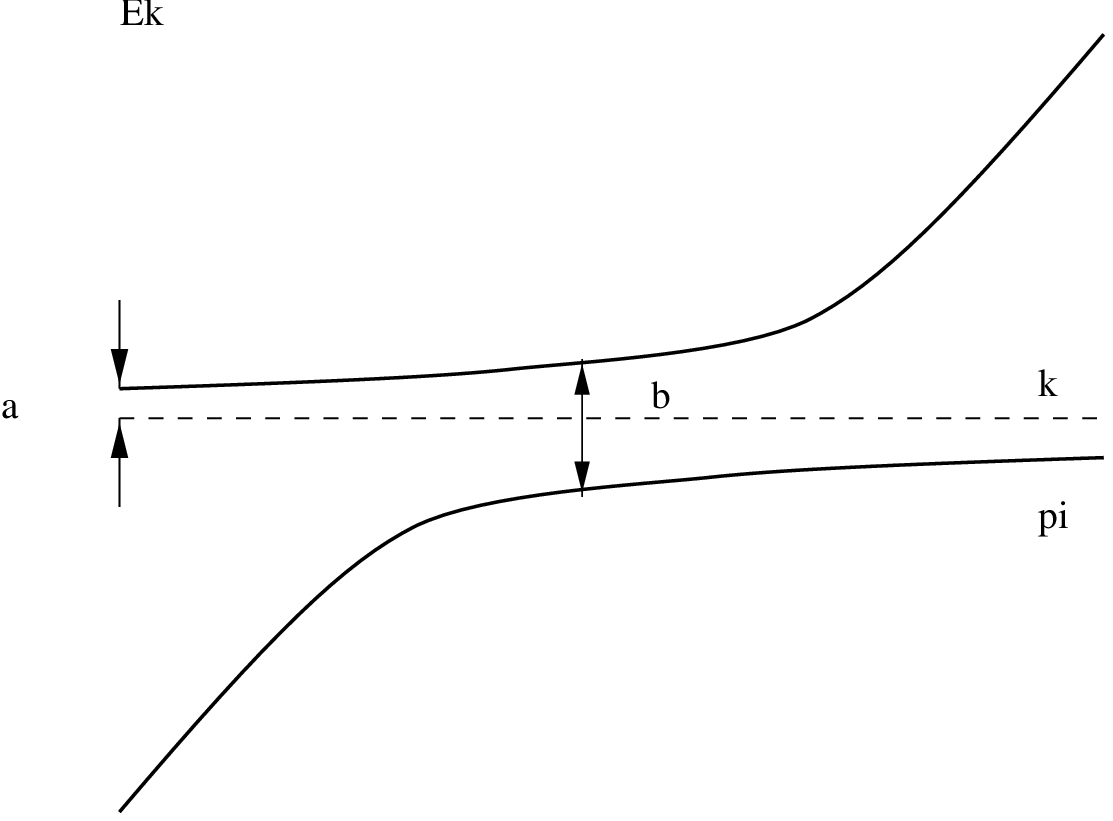}
\end{center}
\caption{Schematic 1D band structure showing the hybridized bands. 
$k$ varies for $k = 0$ to $k = \pi$ and as apparent
charge gap $\Delta_C \sim 2\Delta_{qp}$ is smaller than the optical gap
$\Delta_{opt}$.}
\label{hybrid.fig}
\end{figure}

\begin{figure}
\psfrag{gapspin}{$\Delta_{sp}$}
\begin{center}
\includegraphics[width=10cm]{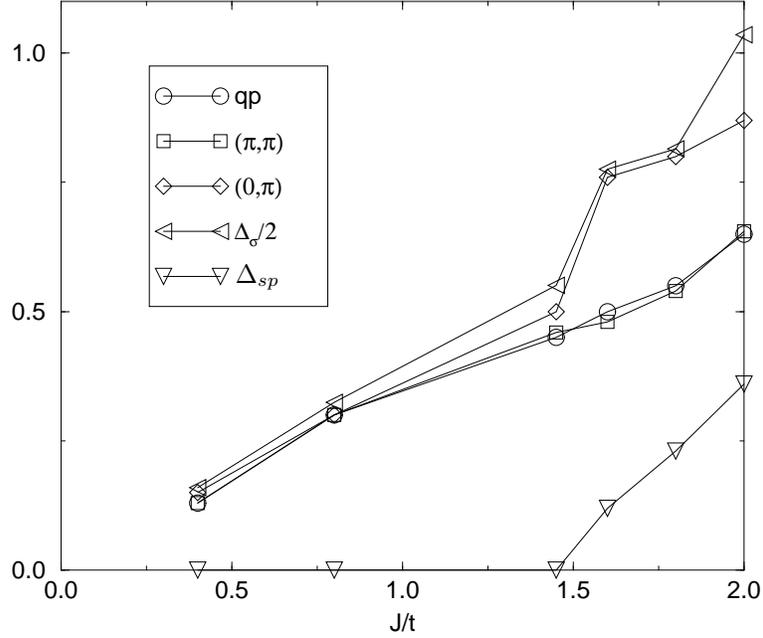}
\end{center}
\caption[]{ Various gaps as a function of $J$. We have considered low enough
temperatures so as to reproduce ground state results and $L=6$. 
The quasi-particle gap is obtained from the DOS, the
gaps at $(\pi,\pi)$ and $(0,0)$ are seen in the spectral functions at those
points, the optical gap, $\Delta_{\sigma}$ 
stems from the optical conductivity and finally, the
spin  gap $\Delta_{sp}$ is taken from Ref. \cite{Assaad99a}. 
As apparent, in the
weak coupling limit, where the quasiparticle dispersion is very flat along the 
$\vec{k} = (\pi,\pi) $ to  $\vec{k} = (0,\pi)$ direction, the optical  and
charge gaps are comparable. (See Fig.~\ref{Akomd.fig}) 
\label{summary.fig} }
\end{figure}

Before considering the characteristic energy scale for the spin degrees
of freedom we comment on the relation between the optical gap - as 
obtained from the low temperature conductivity data - and quasiparticle gap 
(see Sec. \ref{T0.sec}). 
They are not directly related since optical transitions 
involve only zero momentum transfer. 
Starting from the hybridization picture, we can represent the band structure 
as in Fig.~\ref{hybrid.fig}. Generalizing this figure to 2D,  we clearly see that  
the smallest optical  gap is
at $\vec{k} = (\pi,0)$ (or equivalent points) and is larger than the 
charge gap $\Delta_c \simeq 2 \Delta_{qp}$. We recall that the  
quasi-particle
gap corresponds to a transfer from a particle at $k=(\pi,\pi)$ in the
lower band to the chemical potential.  We then expect from that naive
argument $\Delta_{opt} > \Delta_C \simeq 2\Delta_{qp}$. More precisely, we can
relate the optical gap to  the gap at $\vec{k} = (0,\pi)$ as  observed in
Fig.~\ref{summary.fig}. 

\begin{figure}
\psfrag{ylabel}{$J^2 \chi_s$}
\psfrag{TJ2}{$T/J^2$}
\begin{center}
\includegraphics[width=10cm]{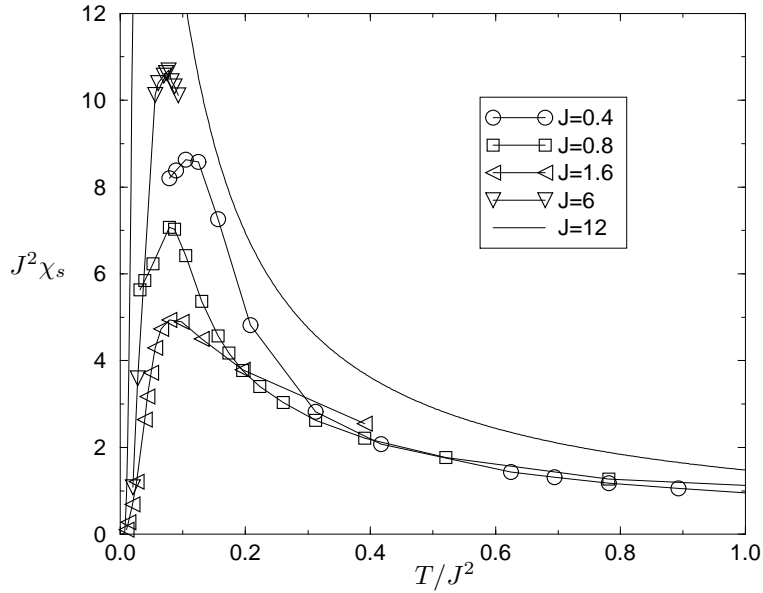}
\end{center}
\caption{Uniform spin susceptibility $J^2\chi_s$ as a function of $T/J^2$ for
various couplings and $L=8$. For $J\le 6$, the maximum
$T_{S}/J^2$ is roughly constant.}
\label{chi0J2.fig}
\end{figure}

To define  a characteristic  energy for the spin 
degrees of freedom, we compute 
the uniform spin susceptibility, $\chi_s= \frac{ \beta}{L^2} 
\left(\langle m_{z}^2\rangle -
\langle m_z \rangle^2 \right) $. Here, 
$m_z  = \sum_{\vec{i}} (n_{\vec{i},\uparrow} - n_{\vec{i},\downarrow}) $ with
$ n_{\vec{i},\sigma} = c^{\dagger}_{\vec{i},\sigma} c_{\vec{i},\sigma}   +
f^{\dagger}_{\vec{i},\sigma} f_{\vec{i},\sigma}$.
In order to observe magnetic properties, it can be necessary to go to very
low temperatures  when $J$ is small. With our algorithm which is free from the sign
problem, we can go down to $T=0.01 t$ for $L=6$ or $T=0.02 t$ for $L=8$. 
\begin{figure}
\psfrag{ylabel}{$J \chi_s$}
\psfrag{TJ}{$T/J$}
\begin{center}
\includegraphics[width=10cm]{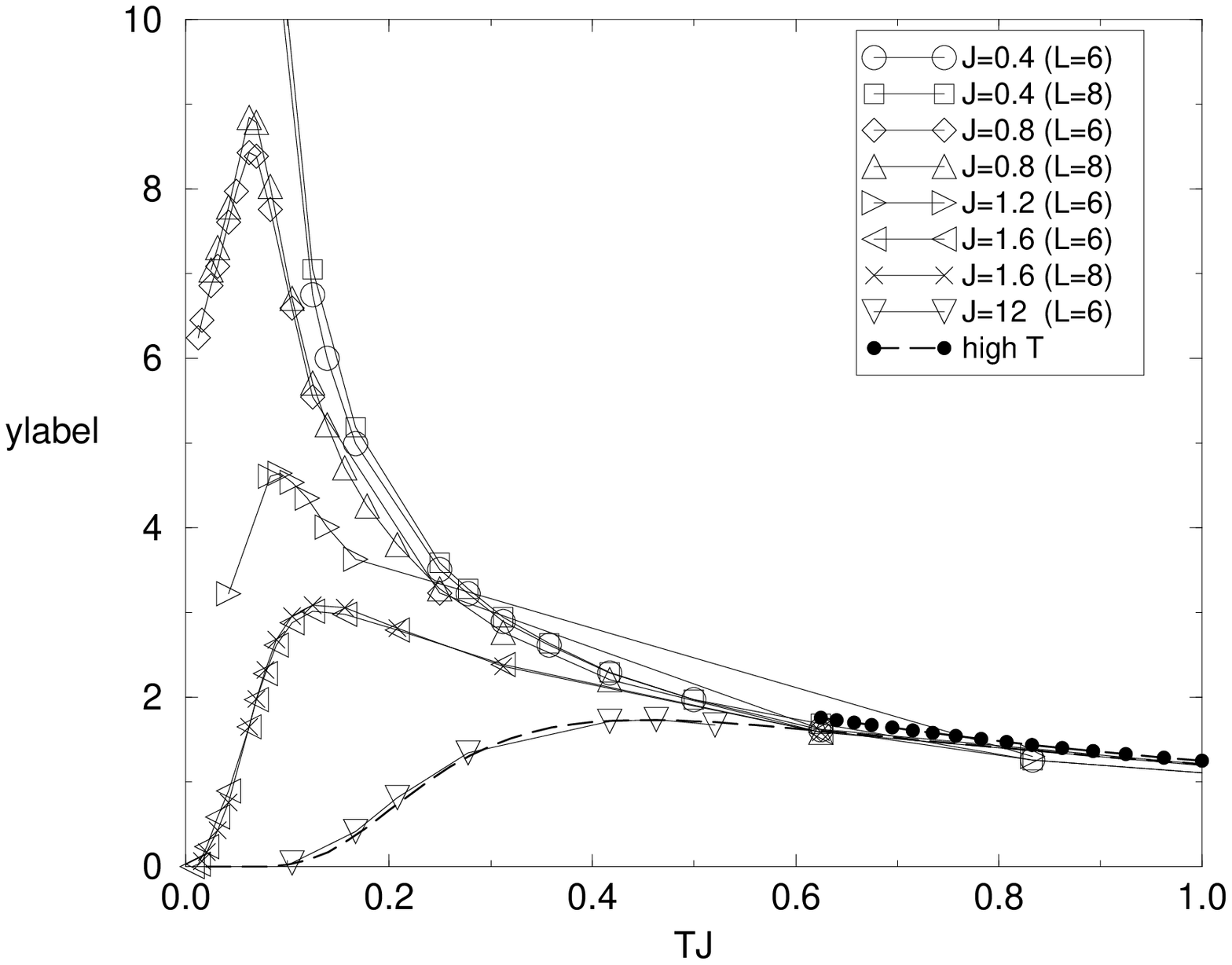}
\end{center}
\caption{Uniform spin susceptibility $J\chi_s$ as a function of $T/J$ for
various couplings and lattice sizes. The high-temperature expansion
(\ref{chis_hiT.eq}) is shown with filled circles  and the large-coupling expansion 
(\ref{chis_largeJ.eq}) is
plotted in dashed line.}
\label{chi0.fig}
\end{figure}
In a high-temperature expansion, $\chi_s$ takes the form:
\begin{equation}
\chi_s=\frac{3}{8T}\left(1-\frac{J}{6T}\right).
\label{chis_hiT.eq}  
\end{equation}
From this expansion, one expects to observe a scaling property
$J\chi_s=f(T/J)$ in this regime. This is indeed what is observed in
Fig.~\ref{chi0.fig} for
$T/J\ge 0.6$. 
We define the magnetic characteristic
temperature $T_{S}$  via the position of the maximum in $\chi_s$.
At large coupling, the physic of the Kondo lattice becomes local. 
In that limit, the susceptibility is easily computed (see a similar
calculation for the charge susceptibility in Eq.~(\ref{chic_largeJ.eq})) and
takes the form~:
\begin{equation}
\chi_s=\beta\frac{1+2 e^{-\beta J/4}}{4+3e^{-\beta J/4}+e^{3\beta J/4}}, 
\label{chis_largeJ.eq}
\end{equation}
which exhibits a maximum at $T_{S}\simeq 0.453 J$. 
In contrast, for smaller $J$, the position of the maximum
clearly increases more slowly than $J$ (see Fig.~\ref{chi0J2.fig}). 
As apparent from  Fig.~\ref{chi0J2.fig} and for the considered values of
$J/t$, 
$T_S$ scales approximately as $J^2$. 

Fig.~\ref{Tmag.fig} plots $T_{S}$ as a function of $J$. In the large
coupling region, we have excellent agreement with the
expansion of Eq.~(\ref{chis_largeJ.eq}); but, for couplings up to $\sim 5$ 
(or bandwidth which is the physical case), $T_{S}$ is well fitted by $\sim J^2$.

The meaning of the energy scale $T_S$ is elucidated by  considering the 
spin susceptibility at the antiferromagnetic wave vector $\vec{Q} = (\pi,\pi)$.
This quantity measures the antiferromagnetic correlation length. 
Indeed, writing the spin-spin correlation functions in space and imaginary time
as $S(\vec{r},\tau)= A \exp(i \vec{r} \cdot \vec{Q} )
\exp(-r/\xi) \exp(-\tau/\xi_{\tau})$, we find that the staggered
susceptibility $\chi_s(\vec{Q})=\int_o^\beta d\tau \int d\vec{r} \exp(-i
\vec{Q} \cdot \vec{r}) S(\vec{r},\tau) \sim \xi^D \xi_{\tau}$ in $D$ dimensions. For the
Heisenberg model~\footnote{Since the charge degrees of freedom are gapped, we
expect that our model is in the same universality class as the O(3) model.},
the dynamical exponent $z$ defined by $\xi_\tau \sim \xi^z$ is equal to 1~\cite{Chakravarty88,Chakravarty89}.
We then
obtain in our case $\chi_s(\vec{Q})\sim \xi^3$.  
\begin{figure}
\noindent
\begin{center}
\includegraphics[width=10cm]{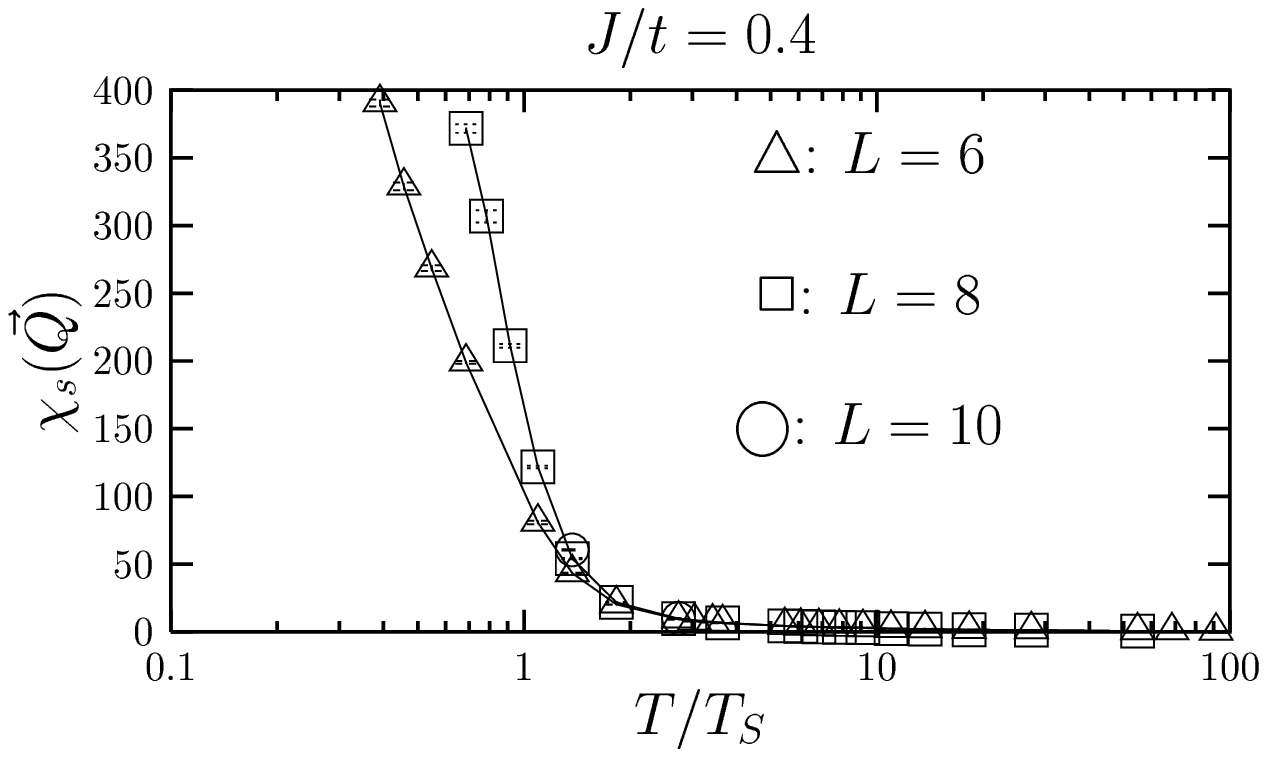}
\includegraphics[width=10cm]{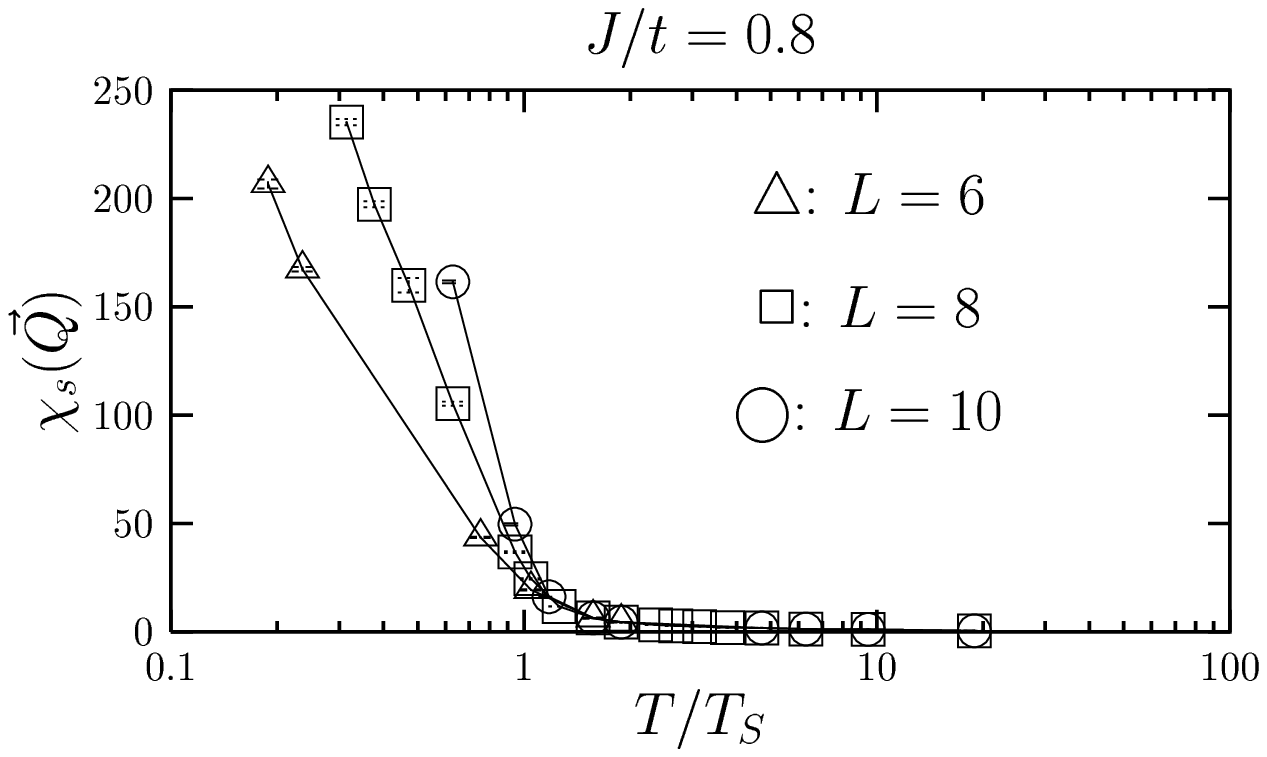}
\includegraphics[width=10cm]{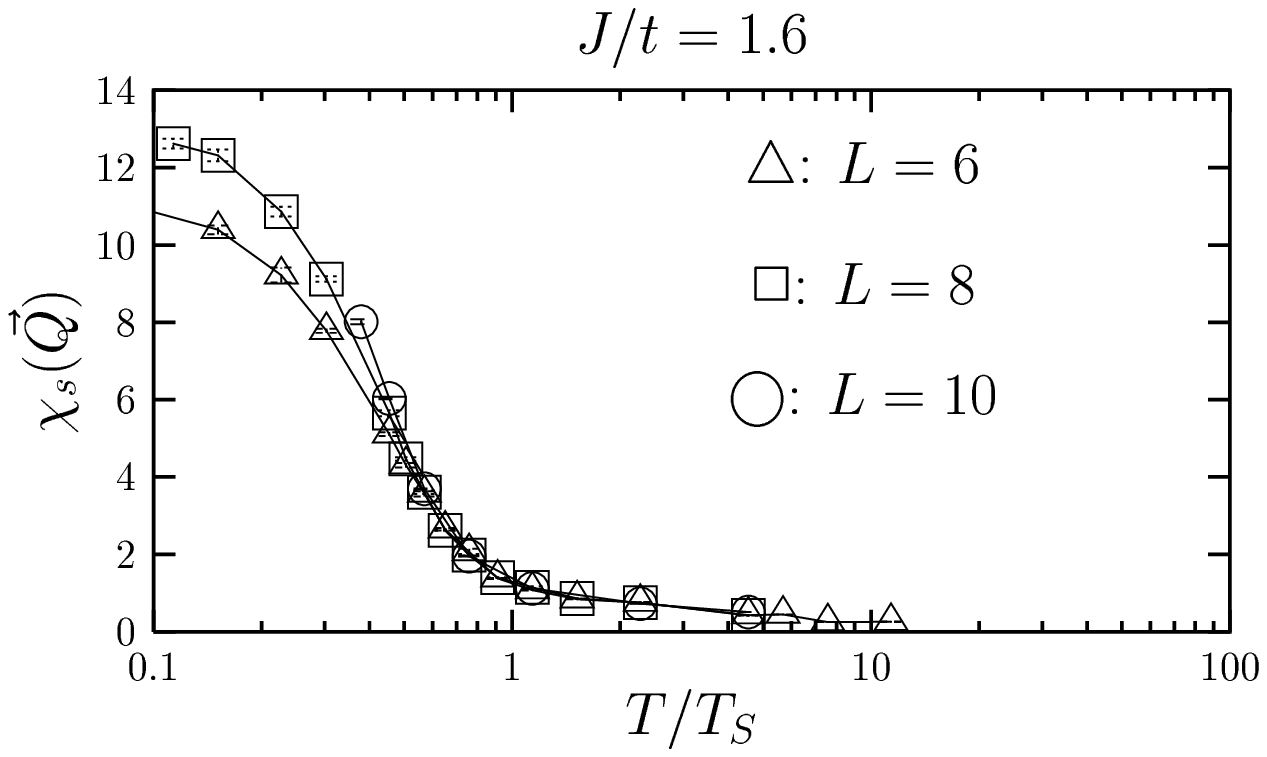}
\end{center}
\caption{Staggered spin susceptibility $\chi_s(\vec{Q})$ for various couplings
and sizes. Since $\chi_s(\vec{Q})\sim \xi^3$, we can extract the behavior of
the antiferromagnetic 
correlation length $\xi$. $T_S \simeq  0.017, 0.05, 0.22 $ for $J/t = 0.4, 0.8, 
1.6$.}
\label{chiq.fig}
\end{figure}

$\chi_s(\vec{Q})$ is plotted in Fig.~\ref{chiq.fig}. As apparent and for the 
considered $J/t$ range,  the  energy scale
$T_S$ marks the onset of {\it short}-range antiferromagnetic fluctuations. At
low temperatures in 
the ordered phase, one expects $\xi$ to grow exponentially as a function 
of decreasing temperature.
On the other hand, in the spin gap phase, $J/t > 1.45$,
the antiferromagnetic correlation length saturates to a 
constant~\cite{Chakravarty89}.

The results of this section are summarized in Fig.~\ref{Tmag.fig}. 
We have defined both a charge, $T_C$, and a spin, $T_S$, scales.
The charge scale  corresponds to the onset of enhanced scattering as a 
function of decreasing temperature due to the screening of the magnetic
impurities.  Apart from  different numerical prefactors $T_C$ scales as $J$ in 
the weak and strong coupling limits. 
From weak to intermediate couplings the spin scale defines the 
onset of short range antiferromagnetic fluctuations and follows a $J^2$ law
in agreement   with the energy scale associated to the RKKY interaction.
At strong couplings, $T_S$ tracks the spin gap.
We note  that we find good  agreement with exact diagonalizations studies 
at finite temperatures  \cite{Prelovsek00}.  This approach is however 
limited to very small cluster sizes  and consequently to high temperatures 
and/or large values of 
$J/t$ where the local approximation becomes valid. Thus those studies cannot
extract the behavior of $T_S$ in the weak coupling limit.

\subsection{Temperature dependence of spectral functions and 
origin of  quasiparticle gap.}
The origin of the quasiparticle gap in the strong coupling limit is the formation
of Kondo singlets. In the weak coupling limit the situation is not
{\it a priori} clear. In the mean-field theory presented in Fig.~\ref{mean_yu.fig}
and retaining only Kondo screening, we obtain an exponentially small gap 
corresponding to the dashed line in Fig \ref{mean_yu.fig}a.  
On the other hand retaining only magnetic ordering, the quasiparticle
gap takes the value $J/4t$ in good agreement with the numerical data.
We note that an exponentially small gap  is  equally obtained
with 
(i) Gutzwiller
approximation~\cite{Ueda86}; (ii) dynamical mean-field
theory \cite{Jarrell95}; (iii) $1/N$
expansion~\cite{Read83}  since those approximations neglect magnetic 
fluctuations. 
In this section, we argue that at or slightly below $T_C$ a small gap 
emerges leading to the quasiparticle dispersion relation 
$ \frac{1}{2}
\left( \varepsilon(\vec{k}) \pm \sqrt{\varepsilon^2(\vec{k}) + \Delta^2 } \right)/2 $
and that the quasiparticle gap of order $J$ is formed only at $T_S$.

\begin{figure}
\psfrag{ylabel}{$N(\omega)$}
\psfrag{omega}{$\omega/t$}
\psfrag{y1}{\scriptsize $0.520$}
\psfrag{y2}{\scriptsize $0.816$}
\psfrag{y5}{\scriptsize $1.83$}
\psfrag{y10}{\scriptsize $0.842$}
\psfrag{y12}{\scriptsize $0.933$}
\psfrag{y15}{\scriptsize $1.23$}
\psfrag{y20}{\scriptsize $1.21$}
\psfrag{y30}{\scriptsize $1.49$}
\psfrag{beta1}{$\beta t=1$}
\psfrag{beta2}{$\beta t=3$}
\psfrag{beta5}{$\beta t=5$}
\psfrag{beta10}{$\beta t=10$}
\psfrag{beta12}{$\beta t=12$}
\psfrag{beta15}{$\beta t=15$}
\psfrag{beta20}{$\beta t=20$}
\psfrag{beta30}{$\beta t=30$}
\begin{center}
\includegraphics[width=10cm]{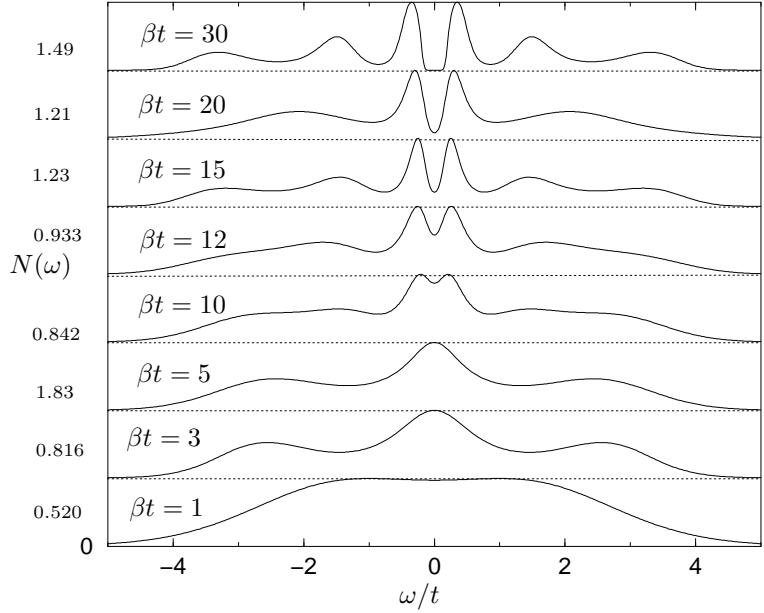}
\end{center}
\caption{Integrated DOS for $J/t=0.8$ on the $L=8$ lattice for various
temperatures shown on the plot. The peak height is normalized to unity and 
normalization factor is listed on the left hand side of the figure.}
\label{dos.fig}
\end{figure}

We start by considering the integrated DOS, $N(\omega)$
obtained with the ME method.  Results are shown in
Fig.~\ref{dos.fig} at $J/t = 0.8$. In the vicinity of the 
charge scale, $T_C = 0.16 t$, one observes a reduction of spectral weight
at the Fermi energy.  Within the mean-field approximation of the KLM 
presented in Eq.~(\ref{H1_KLM}) and (\ref{MFoder}),
this dip in the DOS of the conduction electrons 
follows directly from the occurrence of Kondo screening, i.e. $V \neq 0$.
Hence this feature in $N(\omega)$ at $T_C$ stands in agreement with our 
interpretation of the charge scale $T_C$. As the temperature is lowered
below $T_C$,  the density of states at the Fermi level is further 
depleted and a  gap opens in the low temperature limit.

\begin{figure}
\begin{minipage}{.45\textwidth}
\includegraphics[width=\textwidth]{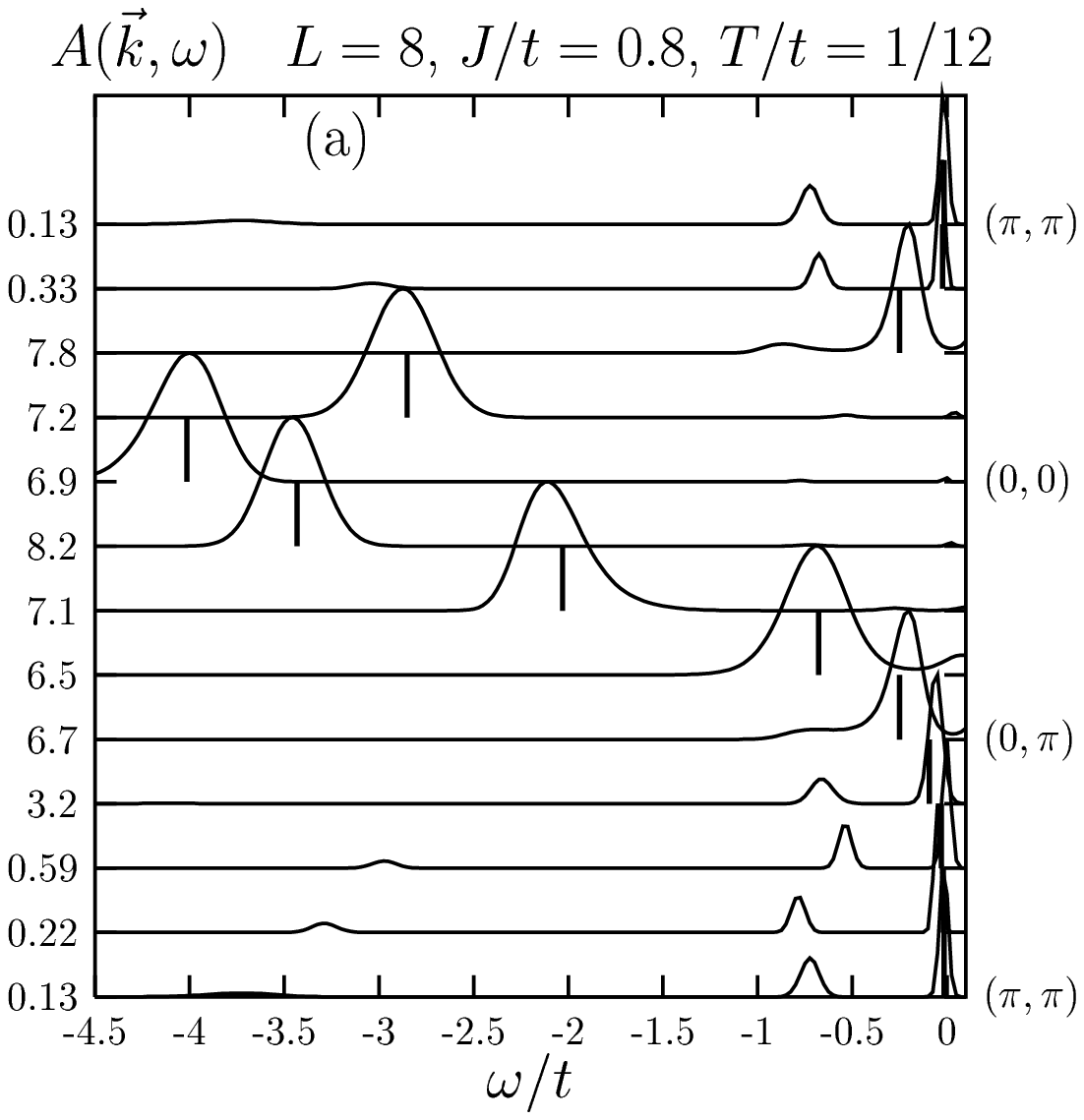}
\end{minipage}
\begin{minipage}{.45\textwidth}
\psfrag{beta20}{\small $\beta t=40$}
\includegraphics[width=\textwidth]{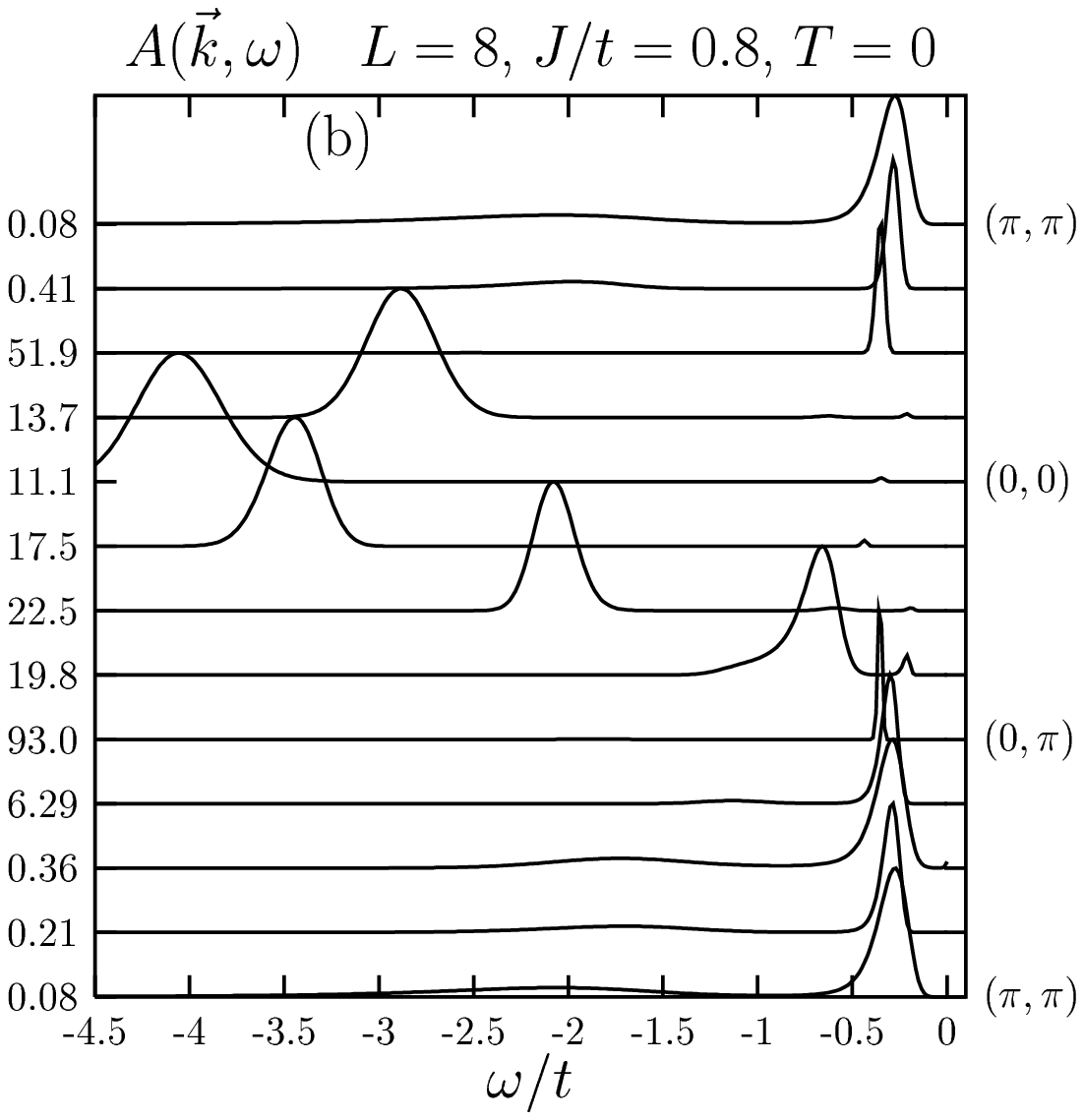}
\end{minipage}
\caption{ (a) Spectral functions  for $J/t=0.8$ and $\beta
t=12$ ($\Delta\tau=0.2$, $L=8$ lattice). Normalization factors  are written on the
vertical axis. At this temperature, $T_S < T < T_C = 0.2 t $, there is a clear
formation of hybridized bands  with quasiparticle gap lying beyond our
resolution. The vertical bars correspond to a fit of the data (see text).
For comparison we have included the $T=0$ results (b).}
\label{Akw.fig}
\end{figure}

In order to gain more insight into the distribution of spectral weight, it
is convenient to compute the momentum-dependent DOS, $A(\vec{k},\omega)$.
The  integrated density of states merely corresponds to the sum 
over all $\vec{k}$ of $A(\vec{k},\omega)$. 
Therefore, we expect the same behavior by decreasing the temperature but we 
have more information on the dispersion relations of the excitations for example. 
Fig.~\ref{Akw.fig}a plots $A(\vec{k},\omega)$ again for $J/t = 0.8$ and at a
temperature $T = 0.083 t$ corresponding to $T_S < T < T_C$. For 
comparison, we have included the $T=0$ data  (see Fig.~\ref{Akw.fig}b) 
As apparent 
the substantial spectral weight of the $\vec{k}$ points on the 
non-interacting Fermi line i.e. $ \vec{k} = (0,\pi),
(\pi/2,\pi/2)$ has shifted to lower energies.  This is the origin of the 
decrease in spectral weight observed at the Fermi level 
in the integrated DOS at $T \simeq T_C$.
However, the flat dispersion relation around $\vec{k} = (\pi,\pi)$ -  with 
significantly less spectral weight - remains pinned
at the Fermi level. 
The dominant  features of the 
quasiparticle dispersion relation  are well reproduced   
by the fit: 
$ \left( \varepsilon(\vec{k}) \pm \sqrt{\varepsilon^2(\vec{k}) + \Delta^2 } \right)/2 $
with $\Delta   = 0.5t$.  This value of $\Delta $ produces a quasiparticle gap
$\Delta_{qp}  = \Delta^2/16 t \simeq 0.016 t $ which lies beyond  our resolution.
As seen in Fig. ~\ref{Akw.fig}b, $ \Delta_{qp} = 0.28 \pm   0.02 $ in the zero 
temperature limit.

\begin{figure}
\psfrag{omega}{$\omega/t$}
\psfrag{beta4}{\small $\beta t=4$}
\psfrag{beta5}{\small $\beta t=5$}
\psfrag{beta8}{\small $\beta t=8$}
\psfrag{beta10}{\small $\beta t=10$}
\psfrag{beta12}{\small $\beta t=12$}
\psfrag{beta15}{\small $\beta t=15$}
\psfrag{beta20}{\small $\beta t=20$}
\begin{minipage}{.45\textwidth}
\includegraphics[width=\textwidth]{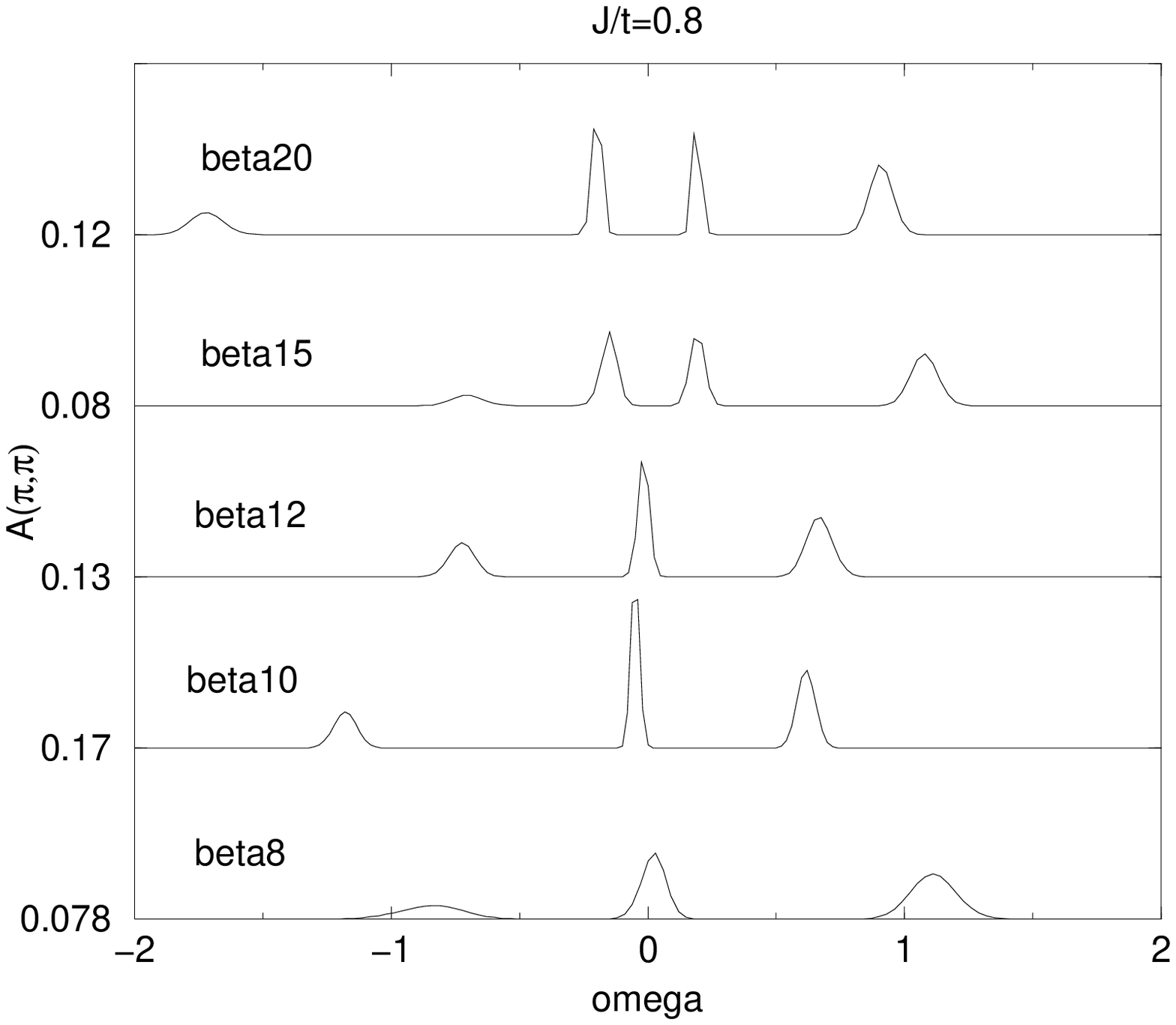}
\end{minipage}
\begin{minipage}{.45\textwidth}
\psfrag{beta20}{\small $\beta t=40$}
\includegraphics[width=\textwidth]{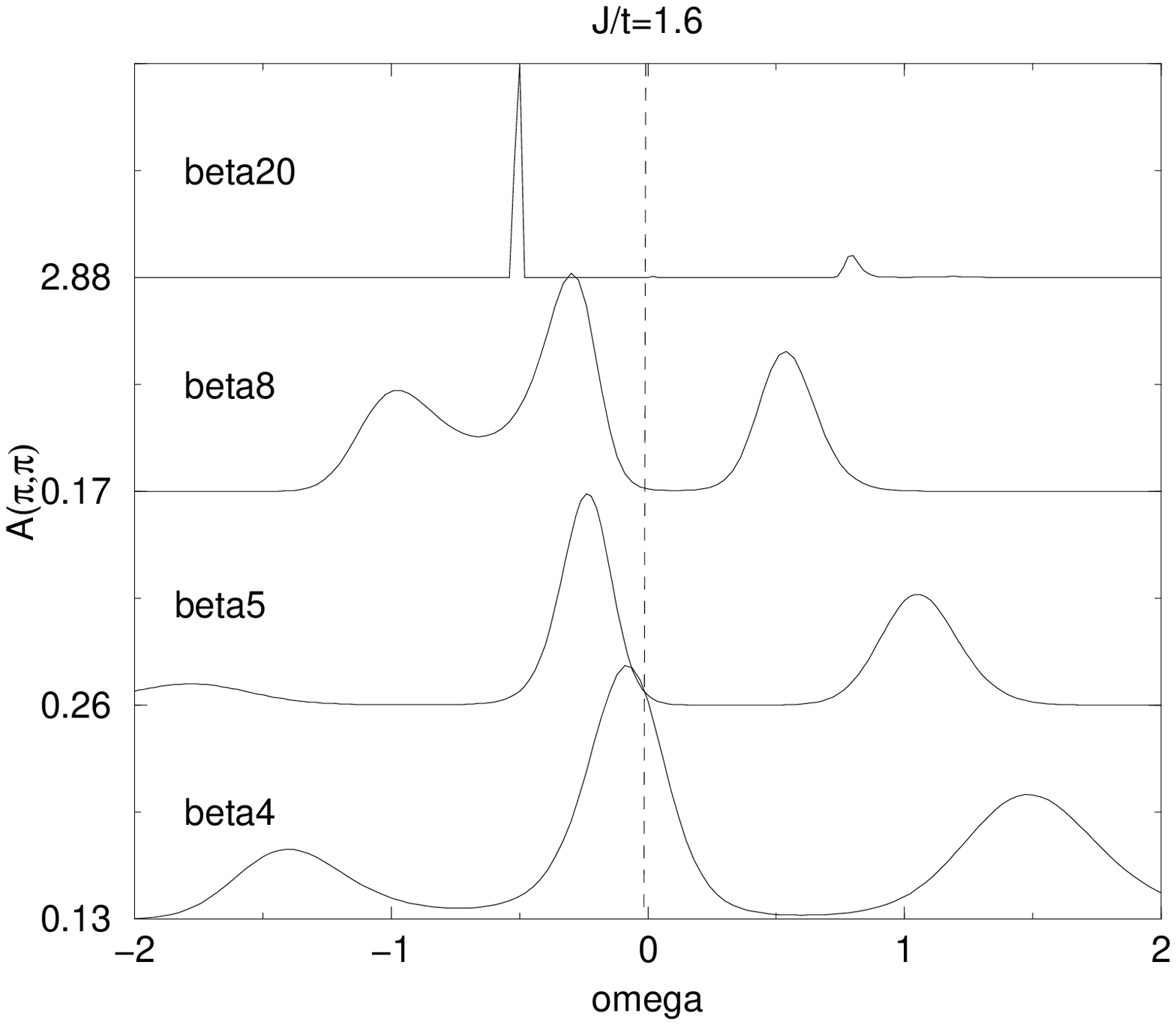}
\end{minipage}
\caption{Spectral functions at $(\pi,\pi)$ for various couplings and
$\beta$ (increasing from down to up) 
($\Delta\tau=0.2$, $L=8$ lattice).  At $J/t = 0.8$, the quasiparticle 
gap of order $J/4t$ opens at a temperature scale comparable to 
$T_S = 0.05$. In the spin gap phase, at $J/t = 1.6$ the quasiparticle gap 
grows smoothly as a function of decreasing temperature.
}
\label{Ak.fig}
\end{figure}

Since the quasiparticle gap is determined by the $\vec{k} = (\pi,\pi)$ wave
vector we concentrate on this $k$-point to analyse the temperature evolution.
As apparent in Fig.~\ref{Ak.fig} at $J/t = 0.8$ the quasiparticle gap of 
order $J$ is
formed approximately at the magnetic scale $T_S = 0.05t$.  Since the 
model is particle-hole symmetric 
$A(\vec{k}, \omega) =  A(\vec{k}+ \vec{Q}, -\omega)$. Thus the fact that the peak
splits symmetrically around the Fermi energy confirms the presence of
shadow bands.
In the spin gap phase the quasiparticle gap originates solely from Kondo
screening. In the mean-field approximation presented in Eq.~(\ref{H1_KLM})  
and (\ref{MFoder}) and
retaining only Kondo screening, the quasiparticle gap will grow 
continuously as a function of decreasing temperatures below the charge
scale. This merely reflects the temperature dependence of the mean-field
order parameter $V$. Precisely this behavior is seen in Fig.~\ref{Ak.fig} 
at $J/t = 1.6$.

\begin{figure}
\begin{center}
\includegraphics[width=10cm,angle=0]{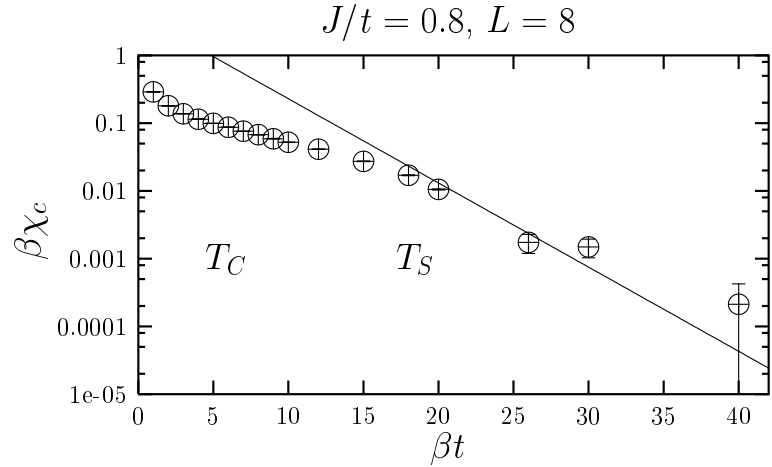}
\end{center}
\caption{ $ \chi_c / \beta $ at $J/t = 0.8$. The solid line corresponds to 
$ \exp(-\Delta_{qp} \beta )$  where quasiparticle corresponds to the value 
obtained with $T=0$ simulations (See Fig.~\ref{Phase.fig}).  Only below the spin
scale, $T_S$, do the data follow the above exponential form. }
\label{Chic_08.fig}
\end{figure}

The evolution of the quasiparticle gap as a function of temperature is equally
seen in the charge susceptibility. At low temperatures one expects $\chi_c =
\beta  \exp(-\Delta_{qp} \beta)$. As apparent from Fig.~\ref{Chic_08.fig},
it is only below $T_S$ that the data follows the above exponential form.

\subsection{Specific heat}
\begin{figure}
\psfrag{chic}{$\chi_c$}
\psfrag{chis}{$\chi_s$}
\psfrag{cv}{$C_v$}
\psfrag{T}{$T$}
\psfrag{JT4}{$J/t=0.4$}
\psfrag{JT8}{$J/t=0.8$}
\psfrag{JT16}{$J/t=1.6$}
\begin{center}
\includegraphics[width=10cm]{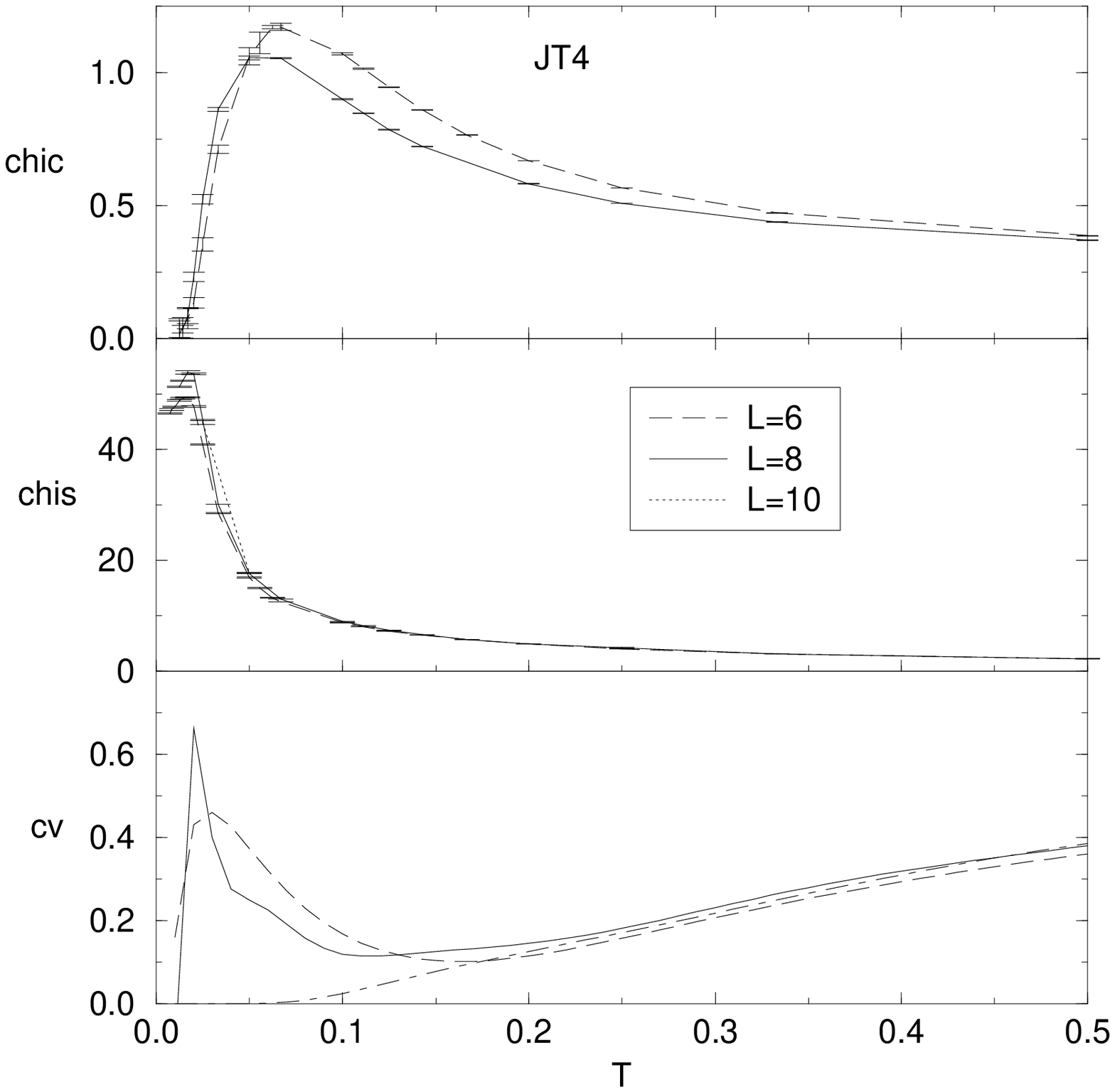}

\includegraphics[width=10cm]{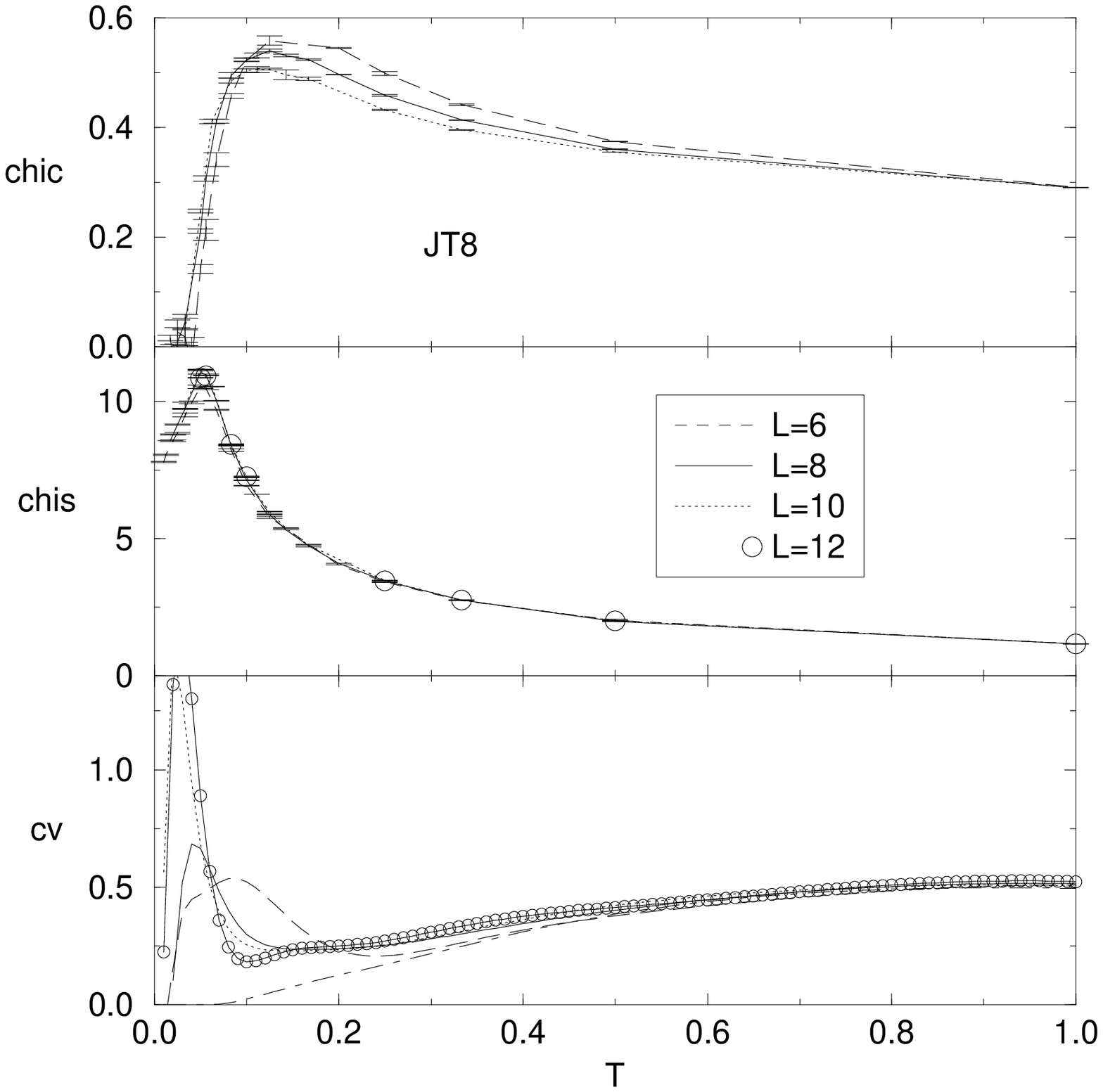}

\includegraphics[width=10cm]{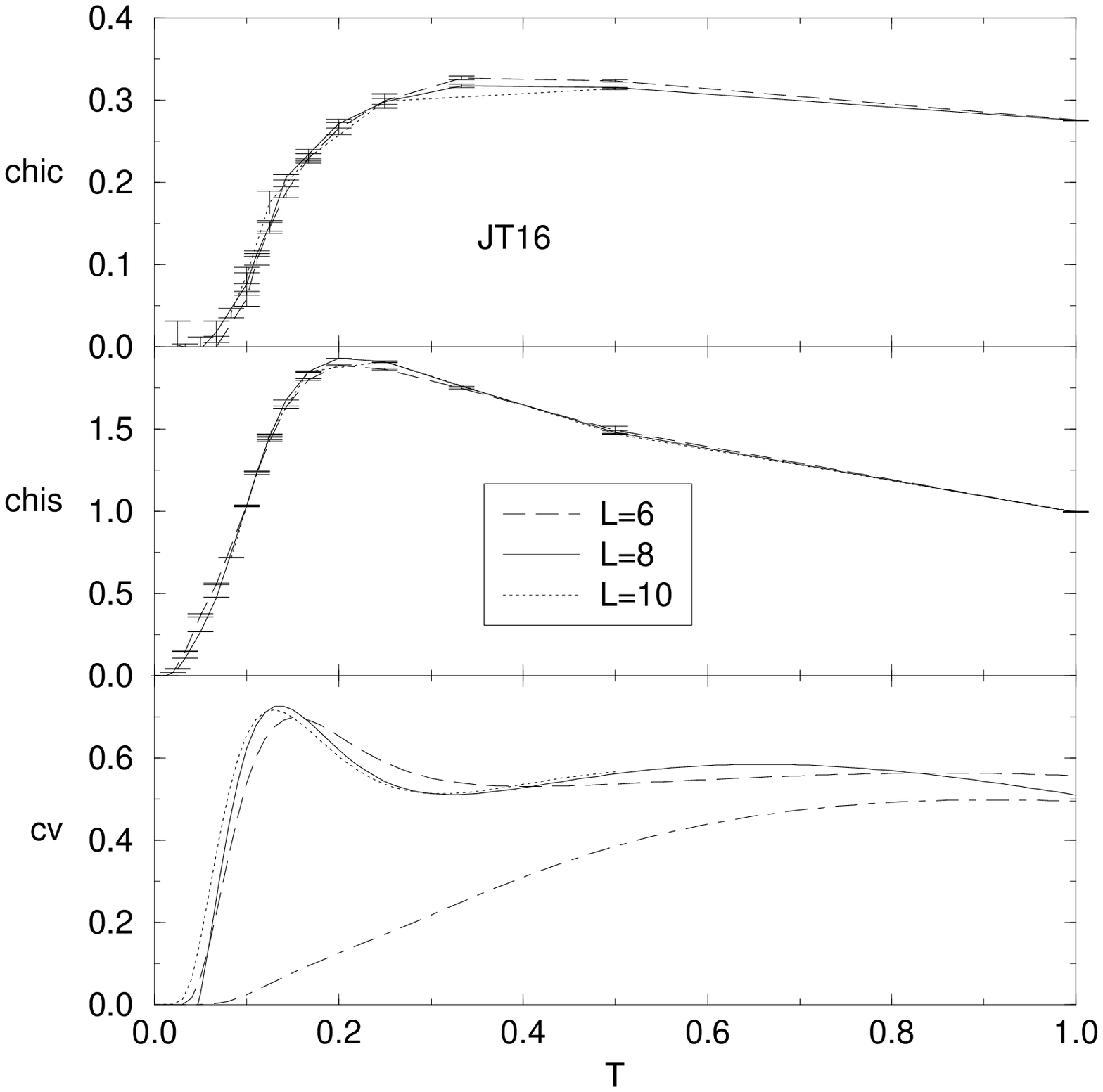}
\end{center}
\caption{Spin, $\chi_s$, and charge, $\chi_c$, susceptibilities as well 
as specific heat, $C_v$, as a function
of temperature for various values of $J/t$. 
The dot-dashed line represents the specific heat of free electrons
on $L=8$.}
\label{thermo.fig} 
\end{figure}

Finally, we consider the specific heat which contains information on both 
spin and charge degrees of freedom.
In principle one can obtain
the specific heat $C_v(T)$ by direct calculation of  the fluctuations
of the internal energy $E(T)$: $C_v(T)=\frac{1}{N} dE/dT=\frac{1}{N}(\langle
H^2\rangle -\langle H\rangle^2)/T^2$. However, this method produces very
poor results at low temperatures.  We have thus  used a ME method to
compute $C_v$ as proposed in~\cite{Huscroft00}. 
In Fig.~\ref{thermo.fig}, we show $C_v(T)$ as well as the uniform spin an charge 
susceptibilities  for various couplings as a function of temperature. 

At $J/t=0$, the specific heat is given by the sum of a delta function at
$T=0$ for the localized spins and the specific heat of free fermions. By
switching on the coupling, they are combined to form a two-peak structure. 
The broad peak at high temperature $T\sim t$ is almost independent of the
coupling $J$ and is rather similar to the free electron gas. The sharp peak at
lower temperatures strongly depends on the exchange constant. It shifts
toward higher temperatures and becomes broader with increasing $J/t$. 
The location of this peak  tracks the magnetic scale $T_S$
indicating that its origin comes from the spin excitations. 
In the spin gapped phase, we note that  the
overall features of $C_v$ agree with the 1D case~\cite{Shibata98}.

\section{Summary and Conclusions}

We have presented a detailed numerical study of  ground state and
thermodynamic properties of the ferromagnetic
and antiferromagnetic half-filled KLM model on a square lattice.
From the technical point of view, we have described and used an efficient 
(i.e. free of the minus-sign problem) auxiliary field QMC method to investigate
the model. Both finite and ground-state algorithms were discussed.
The approach is by no means restricted to the KLM and  may be applied to 
investigate models such as the half-filled
two channel Kondo lattice or various forms of depleted Kondo lattices in 
which the impurity spins are removed in a regular or random way. 
However, we are tied to particle-hole symmetry since only in this case can
we avoid the minus-sign problem.  

In two dimensions the KLM shows a quantum phase transition between 
antiferromagnetically ordered and disordered  states.  This transition
occurs at $J/t = 1.45 \pm 0.05$. The magnon dispersion evolves smoothly 
from its strong coupling form with spin gap at $\vec{Q} = (\pi,\pi)$ to
a spin-wave form in the ordered phase.  The transition may be well
understood in the framework of a  bond-operator mean-field approximation of 
the Kondo necklace model ~\cite{Zhang00}. Here, the disordered phase is 
represented by a 
condensation of singlets with an energy gap at $\vec{Q} = (\pi,\pi)$ for 
magnon excitations. At and below the critical point the spin gap
closes leading to a condensation of both singlets and triplets at the 
antiferromagnetic wave vector.   
The system remains insulating. To a first approximation and as in the 
one-dimensional case, the quasiparticle gap scales a $|J|$  
irrespective of the sign of $J$. 
In contrast, the quasiparticle dispersion relation shows marked 
differences between ferromagnetic and antiferromagnetic couplings. 
For antiferromagnetic couplings the quasiparticle dispersion always has a
structure which follows  the functional form obtained in the non-interacting  
PAM: 
$E_{\pm} (\vec{k}) = \frac{1}{2}
\left( \varepsilon(\vec{k}) \pm \sqrt{ \varepsilon(\vec{k})^2 + \Delta^2} \right) $.
This functional form  is obtained in
various approximations \cite{Zhang00b,Eder98} which take into account
Kondo screening but neglect magnetic ordering. In the antiferromagnetic phase
the above dispersion
relation is merely  supplemented by shadow features. One obtains 
a four band structure which is well reproduced by mean-field theories 
which produce phases with coexistence of magnetic ordering  and
Kondo screening \cite{Zhang00b}. Thus, in the ordered phase screening of the 
impurities is not complete.  The remnant magnetic
moments order due to the RKKY interaction.
Although we cannot dope the system - due to severe minus-sign problems -
it is tempting to assume a rigid band picture and to describe the doped
phase by shifting the chemical potential into the conduction band. Since
the quasiparticle gap is determined by the  $\vec{k} =(\pm \pi,\pm \pi)$ 
points,  the Fermi line will consist of hole pockets around those points
and one expects the Luttinger volume to account both for localized and
conduction electrons. Furthermore, since the band is very flat around those
points a larger effective mass is anticipated. 
Ferromagnetic couplings show a different behavior. In this case, 
Kondo screening is absent but the RKKY interaction present. The  
quasiparticle dispersion is well fitted by the form 
$E_{\pm} (\vec{k}) = \pm \sqrt{ \varepsilon(\vec{k})^2 + \Delta^2} $
corresponding to free electrons subject to an external staggered magnetic 
field. In this case,  again assuming a rigid band picture, 
doping produces a Luttinger
volume containing only the conduction electrons. This contrasting behavior
of the Luttinger volume for the ferromagnetic and antiferromagnetic KLM is 
reproduced in the limit of large dimensions \cite{Matsumoto95}.

From the finite temperature simulations, we can define spin, $T_S$, and
charge, $T_C$, energy scales by locating the maximum in the charge and spin 
susceptibilities.
From weak to intermediate couplings the spin scale follows a $J^2$ law in
agreement with the  energy scale associated with  the RKKY interaction. At  
strong couplings $T_S \propto J$. In contrast both in the weak and strong coupling 
limit $T_C \sim  J$. In the range where $T_S \propto J^2$,
the staggered susceptibility shows a marked increase at $T \sim T_S$. Hence, 
in this range $T_S$ corresponds to the onset of antiferromagnetic fluctuations.
On the other hand, the charge scale determines to a first approximation the 
minimum in the resistivity. Furthermore, at $T_C$ antiferromagnetic intracell
correlations between the $f$- and $c$- electrons are formed and a dip in the
integrated density of states, $N(\omega)$, at the Fermi level is observed. 
Thus, this scale marks the  onset of enhanced 
scattering originating from the 
screening of the magnetic impurities.  In the limit of infinite dimensions, a
similar behavior in the charge degrees of freedom 
is seen but at a much smaller energy scale,
$T_0 \sim e^{-1/2JN(\varepsilon_f)} $ \cite{Jarrell95}.
Apart form a factor $1/2$ in the exponent $T_0$ corresponds to Kondo temperature
of the single impurity problem. In one dimension, a dip in $N(\omega)$ is observed
at an energy scale larger than the spin gap which scales as 
$ e^{-1/\alpha JN(\varepsilon_f)}$ in the weak 
coupling limit (with a numerical estimation of $1 \leq \alpha \leq
5/4$~\cite{Tsunetsugu97_rev} or $\alpha=1.4$ \cite{Shibata99}).
 
In the weak coupling limit, one can analyze the single particle spectral
function at various temperatures. Our results show that the quasiparticle
gap of order $J$ is formed only at the magnetic energy scale. Thus one
can only conclude that the  quasiparticle gap at weak couplings is of magnetic
origin. In contrast at strong coupling, the quasiparticle gap originates
from Kondo screening. The above  stands in agreement with arguments and numerical 
results presented for the one dimensional case \cite{Tsunetsugu97_rev,Shibata99}. 
At weak couplings in 1D, the spin gap becomes exponentially small. Hence,
the time scale associated with  magnetic fluctuations is exponentially 
larger than the time scale relevant for charge fluctuations which is set by $t$. 
The conduction electrons thus effectively feel a  static magnetic 
ordering.  
In 1D and in 2D in the presence of particle-hole symmetry, nesting of the 
non-interacting Fermi surface is present. At a mean-field level and in the
presence of magnetic ordering, this leads to 
a quasiparticle gap $\Delta_{qp} = J/4$. In 2D one may alter the shape of
the non-interacting Fermi
surface so as to avoid nesting by introducing a small nearest neighbor hopping
matrix element. In this case, the mean-field approximation does not produce an 
 insulating state  in the presence of antiferromagnetic ordering. 
Since  nesting is related to particle-hole symmetry, we cannot address this 
question in the QMC approach due to severe sign problems. 
Hence it is worth  paying particular attention to our results at 
weak couplings and $T_C > T > T_S$ before antiferromagnetic correlations set
in.   
Here, Kondo screening
is present but antiferromagnetic correlations  absent.  In this temperature range, 
$A(\vec{k},\omega)$ shows a dispersion relation following that of 
hybridized bands with quasiparticle gap lying beyond our resolution.

We have equally computed the specific heat, $C_v$. This quantity shows a two-{\it peak } structure. The broad high energy ($T \sim t$) 
feature stems from the conduction electrons. The low energy peak 
is very sharp in the ordered phase and tracks $T_S$. 
It is hence of magnetic origin. 

Finally we discuss the relationship of our results to experiments.  Let
us first concentrate on Ce$_3$Bi$_4$Pt$_3$. At $T=100$~K the effective
magnetic moment
of Ce ions starts decreasing \cite{Bucher94}. 
At higher temperatures the Ce ion has a next
to  fully developed moment (i.e. $J=5/2$ as appropriate for Ce$^{3+}$).
At the same temperature scale the real part of the optical conductivity
shows a reduction of  spectral weight  in a frequency range 
of $39 $meV or $450$~K \cite{Bucher94,Degiorgi_rev}.
Those results imply that the opening
of a gap is related to the screening of magnetic impurities and hence, the 
KLM seems to be an adequate prototype model for the description of this class 
of materials.  The above described temperature evolution is precisely  
seen in our numerical 
simulations.  At $T \simeq T_C$ and at {\it weak} couplings,  
the optical conductivity shows a transfer of spectral weight from low frequencies
to frequencies well above $T_C$ (Fig.~\ref{sigma.fig}).  Screening of the
magnetic moments 
start equally at $T \simeq T_C$ (Fig.~\ref{SfSc.fig}).
For the above material, the optical gap is 
estimated by $\Delta_{\sigma} = 39 $meV \cite{Severing91} and photoemission 
experiments suggest a
quasiparticle gap $\Delta_{qp} = 20$meV \cite{Breuer98}. 
At a temperature scale $T \simeq 25$~K 
a gap in the magnetic excitation of $\Delta_{sp} = 12 $meV is observed. Those 
small energy scales imply that small values of $J/t$ should be considered. The
gaps equally satisfy the relation $\Delta_c \simeq 2 \Delta_{qp} > \Delta_{sp}$ as 
obtained in the KLM.   Hence, one should place this material in the parameter
range $ J > J_c$ which in our calculations seem rather large in comparison to the
small charge gap observed in experiments. However, one should keep in mind
that $J_c$ may be sensitive to the properties of the non-interacting Fermi surface.
In particular nesting - which is present in our calculation - will certainly
enhance  the value of $J_c$.  We now turn our attention to CeNiSn.  CeNiSn has a 
transport gap roughly an order of magnitude smaller than  Ce$_3$Bi$_4$Pt$_3$, and
hence - assuming a KLM description 
of the material -  should correspond to smaller values of $J/t$ in comparison 
to  Ce$_3$Bi$_4$Pt$_3$. \footnote{ Recent measurements down to $\sim 0.1$~K 
have found an electronic contribution to the specific heat \cite{Izawa99}. 
This is interpreted in terms of  a finite density of states within the gap.}
This smaller value of $J/t$  leads to signs of
magnetism. Indeed along the $a$-axis of the orthorhombic structure, CeNiSn shows 
a peak in the magnetic susceptibility at $12K$. At the same energy scale, an 
anomaly is seen in the specific heat \cite{Takabatake90}. 
This seems consistent with our results.

We acknowledge useful discussions and communications  with 
R. Eder,
M. Feldbacher,
G. Gr\"uner,
O. Gunnarson,
C. Huscroft, 
Y. Lu,
A. Muramatsu,
H. Tsunetsugu and
G. M. Zhang.
We thank HLRS Stuttgart  for generous allocation of CPU time on the Cray-T3E.


\end{document}